\documentclass[preprint2]{aastex6}

\usepackage{lineno}
\usepackage{graphicx}

\usepackage[utf8]{inputenc}

\begin{document}

\title{Ultraviolet Spectropolarimetry with Polstar: Massive Star Binary Colliding Winds}

\author{Nicole St-Louis}
\affil{D\'epartement de physique, Universit\'e de Montr\'eal, Complexe des Sciences, 1375 Avenue Th\'er\`ese-Lavoie-Roux, Montr\'eal (Qc), H2V 0B3, Canada}

\author{Ken Gayley}
\affil{Department of Physics and Astronomy, University of Iowa, Iowa City, IA, 52242}
\author{D. John Hillier}
\affil{Department of Physics and Astronomy \& Pittsburgh Particle Physics, Astrophysics and Cosmology Center (PITT PACC), \\ \hspace{1cm}  University of Pittsburgh, 3941 O'Hara Street,  Pittsburgh, PA 15260}

\author{Richard Ignace}
\affil{Department of Physics \& Astronomy,
East Tennessee State University,
Johnson City, TN 37614, USA}

\author{C. E. Jones}
\affil{Department of Physics and Astronomy, Western University, London, ON N6A 3K7, Canada}

\author{Alexandre David-Uraz}
\affil{Department of Physics and Astronomy, Howard University, Washington, DC 20059, USA\\
Center for Research and Exploration in Space Science and Technology, and X-ray Astrophysics Laboratory, NASA/GSFC, Greenbelt, MD 20771, USA}

\author{Noel D. Richardson}
\affil{Department of Physics and Astronomy, Embry-Riddle Aeronautical University, 3700 Willow Creek Rd, Prescott, AZ, 86301, USA}

\author{Jorick S. Vink}
\affil{Armagh Observatory and Planetarium, College Hill, BT61 9DG Armagh, Northern Ireland, UK}

\author{Geraldine J. Peters}
\affil{Department of Physics \& Astronomy,
University of Southern California,
Los Angeles, CA 90089, USA}

\author{Jennifer L. Hoffman}
\affil{Department of Physics \& Astronomy,
University of Denver, 2112 E. Wesley Ave., Denver, CO 80208, USA}

\author{Ya\"el Naz\'e}
\affil{GAPHE, Universit\'e de Li\`ege, All\'ee du 6 Ao\^ut 19c (B5C), B-4000 Sart Tilman, Li\`ege, Belgium}

\author{Heloise Stevance} 
\affil{Department of Physics \& Astronomy, University of Auckland, 38 Princes Street, 1010, Auckland, New Zealand}

\author{Tomer Shenar} 
\affil{Anton Pannekoek Institute for Astronomy and Astrophysics, University of Amsterdam, 1090 GE Amsterdam, The Netherlands}

\author{Andrew G. Fullard} 
\affil{Department of Physics \& Astronomy, Michigan State University,567 Wilson Rd., East Lansing, MI 48824, MI, USA}

\author{Jamie R. Lomax}
\affil{Physics Department, United States Naval Academy, 572C Holloway Rd, Annapolis, MD 21402, USA}

\author{Paul A. Scowen}
\affiliation{NASA GSFC, Greenbelt , MD 20771, USA}

\newcommand{\DJH}[1]{{\color{brown} DJH:~#1}}
\newcommand{\commentCEJ}[1]{\textcolor{purple}{{\bf Comment CEJ: #1}}}
\newcommand{\commentYN}[1]{\textcolor{blue}{{\bf Comment YN: #1}}}
\begin{abstract}
The winds of massive
stars are important for both their direct impact on the interstellar medium, and their
influence of the final state prior to supernova.
However, the dynamics of these winds is understood primarily via their illumination from a
single central source.
The Doppler shift seen in resonance lines is a sharp tool for inferring these dynamics, but still
the mapping from that Doppler shift to the radial
distance from the source is ambiguous.
Binary systems can reduce this ambiguity by providing
a second light source at a known radius in the wind, seen from orbitally
modulated directions.
A massive companion also provides unique additional information about wind momentum fluxes,
from the nature of the collision between the winds.
Since massive stars are strong ultraviolet (UV) sources, and UV resonance line opacity in
the wind is strong, UV instruments with high resolution spectroscopic capability are essential
for extracting this dynamical information.
Also, polarimetric capability helps further resolve ambiguities in aspects of the
wind geometry that are not axisymmetric about the line of sight,
because of its unique access to scattering direction information.
We review how the proposed MIDEX-scale mission \textit{Polstar} can use UV spectropolarimetric
observations to critically constrain the physics of colliding winds, and hence radiatively-driven winds in general.  We propose a sample of 20 binary targets, capitalizing on this unique combination of illumination by companion starlight,
and collision with a companion wind, to probe wind attributes over a range in wind strengths.
Of particular interest is the hypothesis that the radial distribution of the wind acceleration
is altered significantly, when the radiative transfer within the winds becomes optically thick
to resonance scattering in multiple overlapping UV lines.
\end{abstract}

\maketitle
\section{Introduction}

Despite comprising the rarest stellar mass group, massive stars ($>$ 8 M$_\odot$) are amongst the most important originators of elements in the Universe because they synthesize and distribute heavy elements, especially oxygen through aluminum, when they explode as supernovae. 
They also form and evolve much faster than lower-mass stars, thereby influencing the formation and
composition of all other stars and their planets.
The drastically stronger stellar wind of a massive star, when compared to the solar wind,
also enriches the interstellar medium, and can lead to significant changes in the star's mass over its lifetime.
This affects its evolutionary path in the H-R diagram, its type of supernovae, whether it becomes a gamma-ray burster, and the compact remnant it produces, along with any gravitational wave signature it might create
if it is a close binary.

Indeed,
most massive stars spend a large fraction of their lives in binary systems with other massive stars, and more than half are thought to engage in mass exchange with a close companion \citep{Sana2012}. 
Tracking these evolutionary effects are the topic of two other \textit{Polstar} proposal objectives,
called \textit{S3} 
\citep{Jones2021} 
and \textit{S4}
\citep{Peters2021};
this white paper relates to objective \textit{S5}, which entails using the proximity of
the two massive stars as a unique probe of the nature of their wind dynamics.

\subsection{Wind Acceleration and Stratification}

Over the past few decades our understanding of radiatively-driven hot, massive-star winds has greatly improved.
But despite this
observational and theoretical progress, fundamental details of radiatively driven winds, such as the velocity structure, still elude us. The so-called {\em $\beta-law$}, although extremely convenient to describe the general density structure of massive-star winds, does not arise from a solid physical description of the winds; it rests on an unlikely simplification that the wind opacity does not
undergo significant change as the wind temperature and density drop with radius, and as the
radiation field shifts from primarily diffusive to primarily radially streaming. 
Attempts to better describe the radial acceleration profile of hot-star winds have been made over the years, such as adjusting the value of the $\beta$ exponent, or in the case of WR stars, adopting a typical value of that exponent in the inner wind, while using in the outer wind a higher
value describing a second extended acceleration regime \citep{Hillier1999,Grafener2005}. 

\begin{figure*}
\centering
\includegraphics[width=\textwidth]{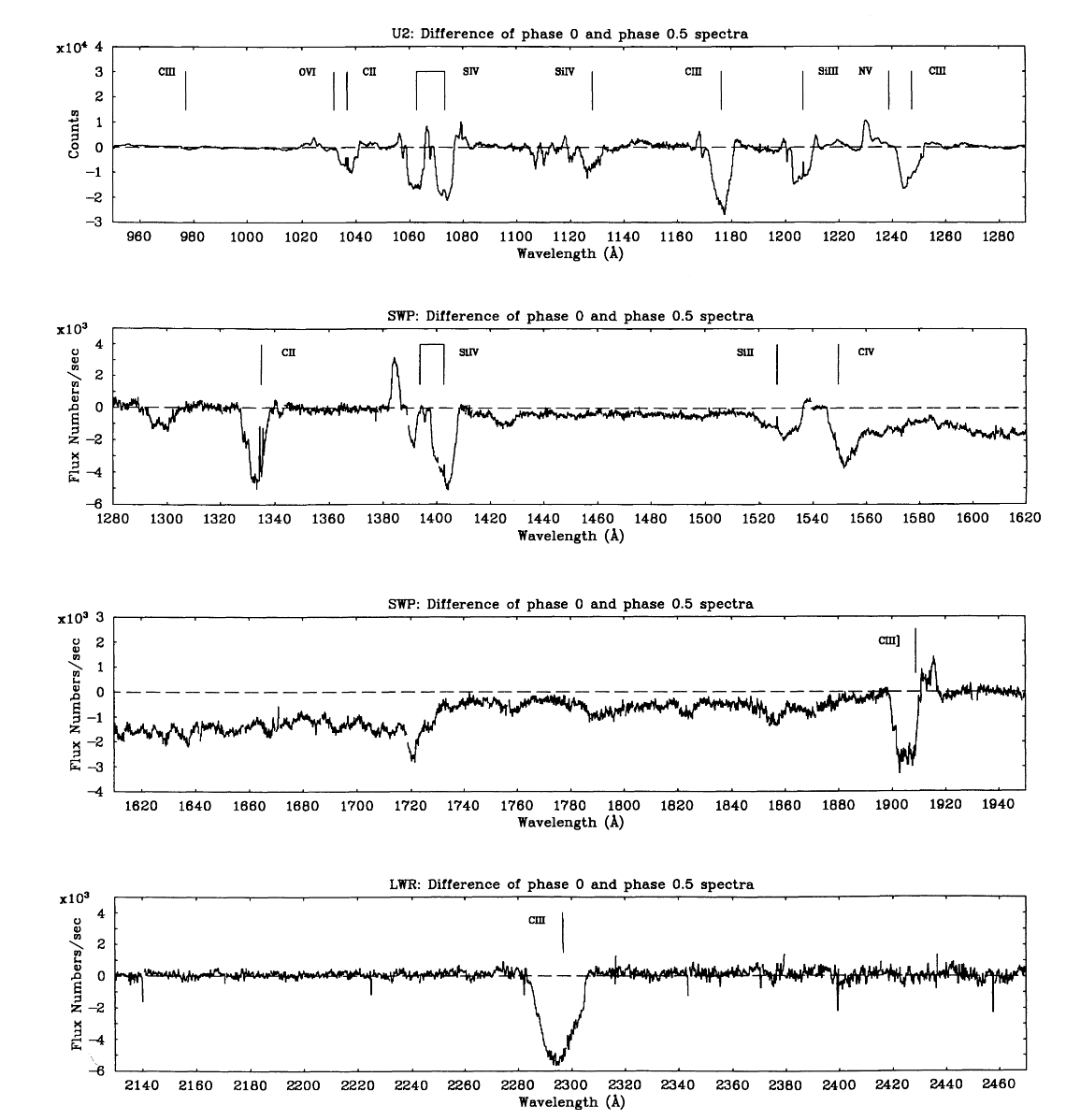}
\caption{Eclipse spectrum of $\gamma$ Velorum. Figure 2 from \citet{1993ApJ...415..298S}}.
\label{figGVdif}
\end{figure*}

The radial dependence of the wind speed has
important consequences, as it controls the steady-state density structure, and creates the diagnostically
significant mapping from observable Doppler shift to radial location.
In optically thick outflows such as those of WR stars, the detailed wind structure is even more important, 
because the hydrostatic surface of the star is hidden from view, leaving the wind as
the only layer of the star that is directly accessible to observation.  
Dense winds also experience additional opacity changes as the temperature drops significantly 
and the radiation field shifts from being highly diffusive in deep layers to more free streaming in outer layers. Therefore, we wish to test the hypothesis that the different character of radiative transport
and opacity gradients in denser winds, compared to less dense winds, changes the nature of the
acceleration as a function of radius.
This white paper describes how this test can be carried out in a special sample of colliding-wind
binary systems.



\begin{figure*}[t]
\includegraphics[width=\textwidth]{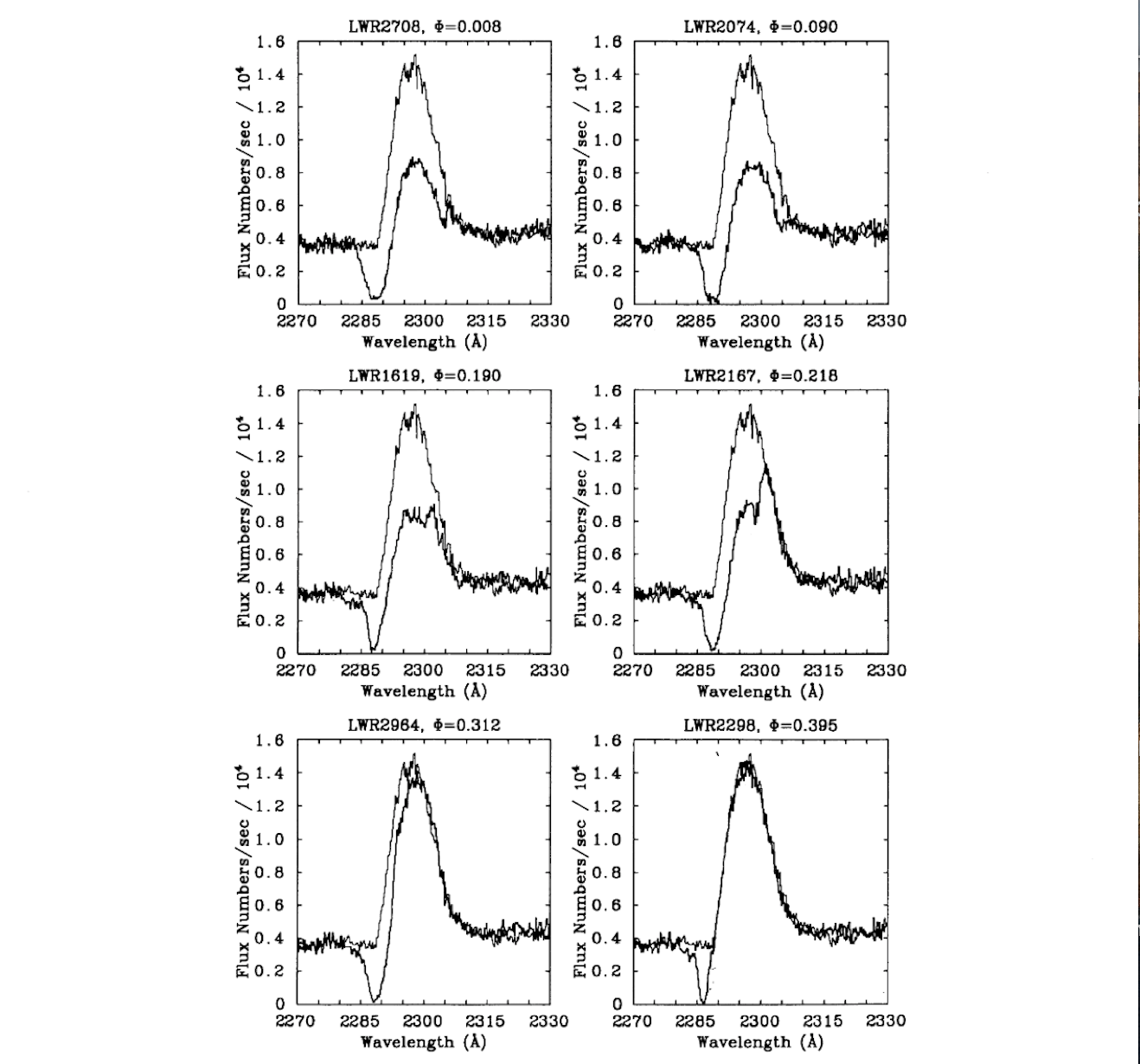}
\caption{ Variations in the C {\sc iii} $\lambda$ 2297 profile with orbital phase. The thin line is the LWR 1316 ($\Phi = 0.534$), and the thick line is the spectrum and phase indicated at the top of each graph. Figure 4 from \citet{1993ApJ...415..298S}.}
\label{figGV}
\end{figure*}

\subsection{Making Use of Massive Stars in Binaries}

Massive stars in binaries offer a unique opportunity to improve our understanding of radiatively driven winds. We can use the light from one star to probe the wind of the other star, allowing us to study the
structure of its wind, which provides information on its density, the velocity law, 
the acceleration zone, and the ionization structure. 
This method has long
been successfully used in the past to investigate, for example, the nature of WR stars
citep{1993ApJ...415..298S}.

An even older version of this approach that focused on light curves \citep{Cherepashchuk1984} 
from the WN5 + O6 V eclipsing binary V444 Cygni, covering the wavelength range from $\sim 2400\,{\rm \AA}$ to 3.5\,$\micron $, demonstrated 
that the WN5 star was hot (90,000 K), compact ($r(\tau_{\scriptstyle}=1)=2.9\,R_\odot$), and had an outflowing atmosphere in which the temperature decreased smoothly with height. 
This helped resolve a long-standing controversy in the nature of Wolf-Rayet stars. 
Additional studies have investigated the stratification in the atmosphere \citep[e.g.,][]{Eaton1985,Brown1986}, and revealed, for example that Fe\,{\sc v} in the wind of the WR star contributed to the atmospheric eclipse signal in the UV.

\cite{Koenigsberger1985} used observations of 6 WR+O systems to show the extended WR atmosphere eclipsed the O stars, and used it to provide constraints on the atmosphere. A further analysis of four binaries by \cite{Koenigsberger1990} showed that the Fe\,{\sc iv}/Fe\,{\sc v} ratio decreased with radius, and that the wind was still being accelerated at 14\,$R_\odot$. \citet{Auer1994} completed a detailed study and showed how line profile variations could be used to investigate the structure of the winds.
Thus, select binary systems have a history of generating breakthrough studies of hot stars, a
spirit that is extended into the present in this white paper.

\subsection{The Diagnostic Potential of Wind Eclipses}

 For a given binary configuration, spectroscopic variations are caused by the absorption of the light from one star by the atoms in the wind of the other.
 Previous UV spectroscopy has shown that absorption occurs mainly in low-excitation and resonance transitions of abundant ions \citep[e.g.][]{Koenigsberger1985}. Therefore, these were called {\em selective wind eclipses}. As shown in Figure \ref{figGVdif}, for the WR+O binary $\gamma$ Velorum \citep{St.-Louis1993}, many ions show a symmetric absorption profile spanning positive and negative velocities in the difference spectrum between phases 0 (WR star in front) and 0.5 (O star in front), as expected. However, note that for certain lines, such as the semi-forbidden C\,{\sc iii}] $\lambda$1909 line and the entire forest of Fe\,{\sc iv} lines, only absorption at negative velocities (towards observer) is observed. This unexpected behaviour still remains to be explained.

Examination of the times series for the same star of lines that behave more as
expected has shown that the absorption profile is widest when the O companion is directly behind the WR star
(called phase 0), and becomes narrower as it orbits to a perpendicular direction with the line-of-sight
(phase 0.5).
This is seen in Figure \ref{figGV}) by considering the difference between the line profile at
the phase shown for the well-isolated C~\textsc{iii} $\lambda$ 2297 line, and that line seen
when the O companion is nearly perpendicular to the WR star.
Here there is evidence of wind eclipse when the O star is seen through the WR wind, at both
red and blue Doppler shifts, spanning the range of projected velocities in the WR wind.

The detailed shape of the eclipse profile in the various transitions will be determined by the opacity encountered by the light from the companion star on its way to the observer and therefore on the projected velocity and ion abundance. As the companion moves through its orbit, different parts of the wind are probed as the line-of-sight differs for different configurations.  This allows to probe different regions of the flow as a function of distance from the center of the star. If the two stars have strong winds, each one can act as a probe for the wind of the other. The combination of all resulting eclipse profiles observed at each position of the orbit contains information on the velocity, density and abundance structures of the wind.

Because of the presence of strong resonance lines of several abundant ions (e.g., due to C\,{\sc iv}, 
Si\,{\sc iv}), far ultraviolet spectroscopy is an ideal tool to study massive-star winds and particularly the spectroscopic variations originating from selective wind eclipse. Emission at any given frequency probes a large volume of the wind where as absorption features primarily probes material along the line of sight to the star. The shape of the continuum light-curve over the binary orbit also allows to probe the column depth of the wind. The opacity of O and WR star winds at optical and UV wavelengths is primarily due to electron opacity.
In WR stars Fe lines in the UV create a pseudo``continuum", and these also contribute significantly to the opacity.

In Figure\, \ref{Fig_rtau} we illustrate the location at which a radial optical depth of 2/3 occurs as a function of wavelength for an O star and a WR star. The minimum radius, at these wavelengths, is set by electron scattering. At other wavelengths we see the influence of strong bound-bound transitions. The Fe forest is particularly prevalent in the WR star. While the actual numerical values are dependent on the
adopted stellar radius, effective temperature and mass-loss rate, the plots do show how different wavelengths have the potential to be used as diagnostic probes in binary star systems.  


{\centering
\begin{figure}
\centering
\includegraphics[width=8.0 cm]{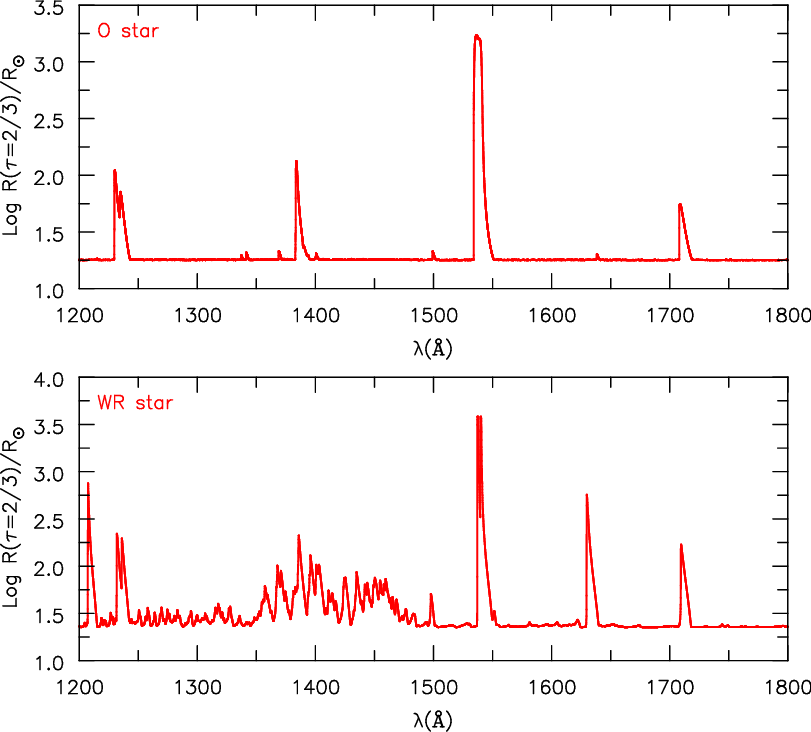}
\caption{Illustration of the location of where a
radial optical depth of $\tau=2/3$ occurs as a
function of wavelength for an O-type and W-R type star.}
\label{Fig_rtau}
\end{figure}
}


\subsection{Colliding Winds in Massive Binaries}

Another aspect of close massive binaries is that the winds of the two stars collide forming a shock-cone structure, generating an asymmetric gas distribution.  Colliding-wind binaries (CWB) can teach us a great deal about the individual stars and their winds because the geometry of the interaction region is strongly dependent on the component winds. Indeed, the details of the geometry and physical characteristics of the shock structure depend on the physical parameters of the individual winds, such as the mass-loss rate and velocity structure, thereby providing a further probe of these parameters. 

Early theories describing these binaries used momentum flux \citep[e.g.][]{Girard1987} and ram-pressure balance \citep[e.g.][]{Kallrath1991}, or hydrodynamic models \citep[e.g.][]{Stevens1992}, until \citet{Canto1996} produced a complete analytical description in the limit of rapid shock cooling (see Section \ref{Canto}). Further work in the area has focused on hydrodynamic simulations \citep[][]{Parkin2008,Lamberts2011,Macleod2020} and predictions of line profiles in both optical \citep{Luehrs1997,Ignace2009_pol,Georgiev2004} and X-rays \citep{Rauw2016_FeXXV,Mossoux2021}. The result of these models, combined with a number of phase-resolved observations \citep[e.g.][]{1990MNRAS.243..662W,1999Natur.398..487T, Rauw2001,Sana2001,Rauw2002,Sana2008,Gosset2009,Parkin2009, 2009MNRAS.395.2221W,2010ApJ...709..632K, Lomax2015,Naze2018,2019MNRAS.488.1282W, Callingham2020}, is that we have a good understanding of the expected geometry of colliding winds.

However, collision of the winds will additionally induce complex geometric structures that must be carefully analyzed to be able to extract the desired diagnostics.  In this context, purely spectroscopic information may not suffice to resolve the potential ambiguities in such a complex structure, including the location and density of the bow shock, the inclination of the orbit, and the potential for asphericity in the intrinsic winds. 

Polarimetric observations are thus a vital tool for revealing details of the mass-loss and mass-transfer structures in these binaries. Both continuum emission (arising from the stars) and line emission (arising from winds, CIRs, or other shock regions) may be polarized in a CWB system via scattering from wind clumps \citep{Harries1998,Davies2005}, accretion disks \citep{Hoffman1998}, jets \citep{Fox1998}, bow shocks \citep{Shrestha2018,Shrestha2021}, and other asymmetric distributions of material or velocities within the system \citep{Schulte-Ladbeck1992,Villar-Sbaffi2005,Fullard2020}. Because scattering near the orbital plane tends to produce repeatable effects over many binary cycles, spectropolarimetric monitoring allows us to reconstruct the 3-D shapes of the regions scattering both continuum and line emission (e.g., \citealt{St.-Louis1993,Lomax2015,Fullard2020}).

Given that massive star winds are strongly ionized, it is reasonable to consider Thomson scattering as the dominant scattering method for photons in the winds, which in turn can polarize the observed light. The resulting polarization is sensitive to the geometry of scattering regions but also depends on the chemical composition of the wind and its  stratification as this will strongly affect the number of available scatterers. Therefore modeling polarimetric observations is a powerful tool to inform us about the geometry and density structures of the colliding winds in interacting systems. This information is critical to our understanding of the roles of binary stars in enriching the ISM and producing explosive transients and gravitational-wave sources.

The classic \citet[][hereafter BME]{Brown1978}  model approximates the time-varying continuum polarization caused by the illumination of stellar winds in a binary system viewed at an arbitrary inclination angle. In this model, the scattering region is described as an optically thin electron gas. The envelope pattern is assumed to co-rotate with the illumination sources, appropriate for steady-state systems in circular orbit. The illuminators consist of one point source at the center of the scattering region and an additional external point source representing the companion. 

\citet{Brown1982} extended the BME model to consider elliptical orbits, and \citet{Fox1994} further extended the formalism to consider finite illuminators. \citet{Fox1994} showed that occultation is only important  in very close binary systems, where separation is less than 10 times the radius of the primary. \citet{Hoffman2003} quantified the polarization produced by external illumination of a disk in a binary system. However, none of these enhancements to the theory included the effects of colliding winds in the time-dependent polarization results. \citet{St.-Louis1993} and \citet{Kurosawa2002} modeled the polarization variations in the colliding-wind system V444 Cyg.  Also, spectroscopic analysis of wind eclipses in general and for this system in particular have been carried out \citep[][]{1994ApJ...436..859A}, but the results have not been compared across a range of binary parameters to extract the important trends.  \citet{HarriesBablerFox2000} used a Monte Carlo formalism to estimate the polarisation caused by the scattering off dust and electrons in a colliding-wind shock cone if it were present in the long-period binary WR137 and found that for such a wide binary, the level would be very small (0.03\%\,). Notwithstanding these individual efforts,  a general formalism has not yet been produced. 

The \textit{Polstar} team includes several experts in polarimetric modeling who will focus on developing the necessary tools to derive physical results from spectropolarimetric signatures. Polarization simulations of bow shocks, which polarize light both when illuminated by a central source and when the light originates within the shock itself \citep{Shrestha2018,Shrestha2021}, can be incorporated into binary models, as we discuss below. We will also investigate the effects of the wind collision regions and the distributed nature of line emission on the wavelength dependence of polarization. Given that complex continuum and line polarization signals associated with CWBs have been observed in several systems \citep{St.-Louis1993,Lomax2015,Fullard2020} and are likely to occur in the UV as well \citep{Schulte-Ladbeck1992,Hoffman1998}, such modeling efforts are critically important. 

We intend to use \textit{Polstar} to secure extensive UV spectropolarimetric timeseries of key massive binary systems in crucial lines with observations well-distributed around the orbital cycle.  Having multiple diagnostics is important -- different diagnostics probe different regions of the flow and trace different physics. For example, P~Cygni absorption (blue shifted absorption associated with a redshifted emission profile) only samples material along our line of sight. Conversely, the emission line  samples a much large volume. Continuum polarization, to some extent, is influenced by  the whole volume. Optical studies do not have the necessary lines for the studies proposed here. A well-distributed  time series is also crucial, since spectra taken at different times probe distinct orientations
of the binary system. The combination of polarimetry and UV spectroscopy, almost unexplored, offers a unique opportunity to resolve ambiguities left by either approach alone.  For example, the rate of change of the polarization position angle caused by the colliding wind interaction is a diagnostic of the orbital inclination that is complementary to light-curve considerations.


\subsection{Small and Large-Scale Structure in Radiatively-Driven Winds}

Growing observational evidence suggests that
radiatively driven outflows are highly clumped, including the
telltale presence of small sub-peaks observed to move from the center to the edges of strong emission lines in both O stars \citep[e.g.,][]{Eversberg1998} and Wolf-Rayet (WR) stars \cite[e.g.,][]{Lepine1999}.  Inconsistencies in line strengths \citep[e.g.,][]{Hillier2003,Crowther2002,Massa2003,Fullerton2006} and the relative strength of electron scattering wings to their adjacent profiles \citep[e.g.][]{Hillier1991,Hamann1998} also indicate that the winds are clumped. Large-scale structures are also thought to be present in the winds of most, if not all, OB stars, as shown by the presence of NACs and DACs in their ultraviolet spectra \citep[e.g.,][]{Howarth1989,Kaper1996} that are thought to be the observational signature of Corotating Interaction Regions \citep[CIRs,][]{1996ApJ...462..469C, Mullan1986}.

These important structural effects are the subject of a different white paper \citep{Gayley2021},
but can be further elucidated by the special diagnostics available in colliding-wind binaries.
Small and large-scale wind structures change the density structure of the winds, and this influences the nature of the wind acceleration being probed in this white paper.  
Further, winds may be porous both spatially (porosity) and in velocity space \citep[called {\em vorosity};][]{Oskinova2007, Owocki2008}. Spatial porosity affects both line and continuum radiation transfer whereas vorosity, which refers to gaps in velocity space, only effects line transfer. These phenomena lead to uncertainties in derived parameters such as mass-loss rates, luminosities, radii, and abundances. In turn, this limits our ability to place accurate constraints on stellar and galactic evolution.
As will be seen below, two very bright systems ($\gamma$ Velorum and $\delta$ Ori)
that will be targeted for close examination for 
clumping effects are also targeted here by virtue of their binary status, and these systems will
be especially informative when the \textit{Polstar} diagnostics utilized in the two objectives are combined.

\subsection{Mass-loss rates}

Mass-loss rates included in stellar evolution calculations typically use a fitting law with an adjustable parameter \citep[e.g. MESA or Geneva code, ][]{Paxton2011, Meynet2015}. These laws are based on a combination of theoretical and observational studies of mass loss in O stars.  For radiatively-driven winds of hot stars, the theoretical mass-loss rates of \cite{Vink2001} and \cite{Nugis2000} are typically adopted. For more luminous stars, they are in rough agreement (factor of 2) with empirically derived mass-loss rates, but for late O stars the rates can differ by an order of magnitude.
Even factor of 2 uncertainty is a significant problem for understanding how massive stars evolve
and effect their environment.

Observational mass-loss rates are generally derived using an empirical velocity law, and by necessity one is 
forced to make assumptions regarding the inhomogeneity of the wind \cite[e.g.,][]{Hillier2003,Bouret2012,Sander2012, Shenar2015}. Typically, the wind is assumed to be clumped, with a fraction $f$ occupied by the clumps. Porosity is generally ignored, and the interclump medium is assumed to be void. To first order, empirical modeling derives $\dot M/\sqrt{f}$.  Constraints on $f$ can be placed by using the strength of electron scattering wings in WR stars, and by using P Cygni profiles in O stars. The latter, however, is more uncertain because of the effects of porosity and vorosity.

Current radiative transfer codes can, potentially, derive the velocity law and mass-loss rates from first principles \cite[e.g.,][]{Vink2001,Sander2020,Bjorklund2021,Vink2021}. However such codes make assumptions about clumping, and may use approximate radiation transfer techniques such as the use of the Sobolev approximation \citep{Sobolev1960}. Further, there are still uncertainties about the nature and role of microturbulence at the sonic point
\citep{Hillier2003,Lucy2007}, and the role of inflation, convection and density inhomogeneities in optically thick regions of the flow \citep[e.g.,][]{Ishii1999,Petrovic2006,Cantiello2009,Grafener2012,Grassitelli2016}. Finally there are uncertainties in the atomic data. 

As a consequence of these uncertainties, spectra derived from models constructed on first principle generally do not provide a good fit to observed spectra. At best,  mass-loss rates must be regarded as uncertain to at least a factor of two, and in some cases the uncertainty is much larger. However, a factor of two variation in mass-loss rate can have a strong influence on a star's evolution. This is seen in theoretical calculations \citep[e.g.,][]{Maeder1981}, and is likely illustrated by the different massive star populations in the LMC and Galaxy, as line-driven mass-loss rates depend strongly on metallicity \citep[e.g.][]{Vink2001}.
Accurate mass-loss rates require an understanding of the density structure, which connects to the
velocity structure and the physics of wind driving.
Thus the objective considered in this white paper is also relevant to establishing more accurate
mass-loss rates in specially selected systems, 
using the unique diagnostics available in these binary systems.

\section{PolStar Capabilities and Data Products}

Using \textit{Polstar} \citep{2021SPIE11819E..08S}, we intend to probe the details of the wind density of the stars in these systems by using the light from the other star as a probe.  Because the presence of two stars with strong winds in a binary generally leads to the formation of a shock-cone structure, we must also take this into account and actually we intend to use it to further constrain the physical characteristics of the outflows. Finally, companion stars are also not independent light sources -- their radiation field can modify the wind of the other star, and in close binaries one or both winds may not reach terminal speed either because of the collision or because of the companion's radiation field. Radiative braking, the deceleration of one wind by the radiation field of the secondary, will modify the location and structure of the bow shock. The combination of UV spectroscopic time series, polarimetric observations, and spectra modeling will ultimately allow us to constrain the mass-loss rates to unprecedented accuracy.



Phase-dependent observations of systems with elliptical orbits or with different separations will allow us to investigate how the radiation field of one star influences the wind of the second star. Our strategy to address these questions is to study a wide range of wind densities in binary systems with high enough inclination and short enough periods for the companion light to be seen deeply through the primary wind. For the brighter stars, we intend to obtain a full phase coverage of ultraviolet spectropolarimetric observations for each system at the highest possible spectral resolution.  Depending on the brightness of a given target, we can bin the spectral resolution as necessary to secure sufficiently high quality signal-to-noise to identify telltale spectral features and polarization variability, with good phase coverage.

We have identified a sample of 20 massive binary systems for which there is evidence in the literature of the presence of a colliding wind structure, and which can be observed with Polstar. 
These are listed in Table 1.

\begin{table*}[]
\caption{Target list }
    \centering
        \begin{tabular}{l                         |c|c|c|c|c|c|c}
     \hline\hline   Star & HD number & Type & Period (d) & V & UV Flux* &Ch. 1$\dagger$  &Ch.2$\dagger$ \\
        &&&&&&exp time(s)&exp time(s)\\
        \hline
        \hline
{\bf Ch.1 spec., N=1}&&&&&&&\\
{\bf and polar. (SNR=3000)}&&&&&&&\\
Gamma Velorum & 68273 & WC8+O7.5III-V & 78.05 & 1.83 & 263 & 2000 & 60\\
Delta Ori & 36486 & O9.5II+BO5.III & 5.7 & 2.41 & 147 & 3550& 60\\
29 CMa & 57060 & O7Iaf+O9 & 4.4 & 4.95 & 7.5 & 70200& 60 \\ 
\hline
{\bf Ch.1 spec., N=1 (SNR=100)}&&&&&&&\\
{\bf Ch. 2 polar. (SNR=3000)}&&&&&&&\\
Plaskett & 47129 & O7.5I+O6I & 14.4 &  6.06 & 1 & 600 & 1840\\ 
WR22 & 92740 & WN7h+O9III-V &80 & 7.16 & 1.1 & 510&1680\\
   &152248 & O7.5III(f)+O7III(f) & 5.8 & 6.05 & 0.8 & 720 &2300\\
WR79 & 152270 & WC7+O5-8 & 8.89 & 6.95 & 0.5 &1150&3680\\ 
& 149404 & O7.5If+ON9.7I & 9.8 & 5.52 & 0.5 & 1150&3680\\
\hline
{\bf Ch.1 spec., N=1.5 (SNR=100)}&&&&&&&\\
{\bf Ch. 2 polar. (SNR=3000)}&&&&&&&\\
WR133 & 190918 & WN5o+O9I & 112.8& 6.7 & 0.35 & 1110&585\\
WR42 & 97142 & WC7+O5-8 & 7.886 & 8.25 & 0.22 & 1800 &930\\
Eta Car(2020) & 93308B & LBV & 2022.7 & 6.21 & 0.2 & 1950&1050\\
WR 25 & 93162 & O2.5If*/WN6+O & 208 & 8.84 &  0.13&4020&1590\\
WR140 & 193793 & WC7pd + O5.5fc & 2895.0  & 6.85 & 0.1 &4020 & 2100\\
& 93205 &O3.5 V((f)) + O8 V &6.0803 &7.75&0.1 & 6300&2100\\
\hline
{\bf Ch.1 spec., N=3 (SNR=100)}&&&&&&&\\
{\bf Ch. 2 polar. (SNR=1000)}&&&&&&&\\
WR137 & 192641 & WC7+O9e & 4766 & 8.15 & 0.04 & 5100 & 5100\\
HD5980 & 5980 & WN4+O7I & 19.3 & 11.31 & 0.04 & 5100 & 5100\\
V444 Cygni & 193576 & WN5+O6II-V & 4.21 & 8 & 0.02 & 11100 & 11000\\
WR69 & 136488 & WC9d+OB & 2.293 &9.43& 0.015 &15600&13680\\
WR127 & 186943 & WN3b+O9.5V & 9.555 & 10.33 & 0.012&21400&17160\\
WR21 & 90657 & WN5o+O4-6 & 8.25443 & 9,76 & 0.011&22800&18720\\
\hline
\end{tabular}
   *UV Flux at 1550A ($10^{-10}$ erg/s/cm$^2$/\AA)\\
$\dagger$ at 1500\AA, D=60 cm,  R=30000/N for spectroscopy
    \label{tab:target_list}
\end{table*}


\subsection{Spectroscopic Experimental Design}

In order to monitor the spectroscopic variability caused by selective wind eclipses and the presence of a colliding wind shock cone and to capture all the subtleties of the phase-related changes, we will use Polstar to secure a series of 10 high-resolution spectra, well-distributed over the orbital phase of each binary on our target list. We plan to observe the 8 brightest systems at the highest resolution (R=30000), the 6 next bright binaries at a slightly lower resolution (R=20000) and finally the 6 less UV bright targets at a more moderate resolution (R=10000). Including fainter targets broadens our system parameters and allows
us to include especially interesting targets with overlap with other \textit{Polstar} objectives,
such as V444 Cygni, a well-studied WR eclipsing binary, HD5980, for which one component star has recently undergone an LBV-type eruption, WR 69, a system including a WC9 type star that produces dust, and WR 137,
a rare system consisting of a WR star and an Oe star (with a decretion disk) in a wide orbit. 
For the three brightest targets, we require a signal-to-noise ratio (SNR) of 3000 to reach the necessary
high polarimetric precision, while also achieving the
high spectral resolution needed to detect the details of the selective wind absorption profile, and details of the clumpy nature of the winds and of the colliding-wind shock-cone. For the other systems, we aim for a SNR of 100 (already 5-6 times higher that what is currently available from IUE data), which will still provide excellent quality data. We plan to use the knowledge gained with our test systems to enhance the interpretation of the lower SNR observations.
\
\ 
\subsection{Polarimetric Experimental Design}

We have selected three massive binaries that are bright in the ultraviolet to observe at the very highest spectroscopic resolution in spectropolarimetry to serve as test cases. These include hot, massive stars having reached different stages of evolution from bright giants to WR type. 
Although our standard polarization precision for this experiment denoted \textit{S5} is
$1 \times 10^{-3}$, for these unusually bright systems, 
we aim for a SNR of 3000 in order to be able to detect the very faint polarization signals from the colliding wind shock-cone,
down to a precision of $5 \times 10^{-4}$. 
We also plan to observe these targets in channel 2 at lower spectral resolution, and appropriately shorter
exposures, in order to characterize the wavelength dependency of the polarization at similar precision. 
For the other binaries, our strategy is to obtain the highest SNR possible  for the UV flux of the target (either 3000 or 1000) but at lower spectral resolution using observations in channel 2. Channel 1 observations at a lower SNR are also important to characterize the wavelength dependency of the polarization.  For all polarization measurements of our targets, we will be able to remove the non time-variable but wavelength-dependent interstellar polarization vector from the interstellar medium between each target and the observer, using the method described in \citet{2021arXiv211108079A}.
\
\vskip 1.5truecm

\section{Polarisation modeling}\label{sec:pol}
\subsection{The Bow Shock Opening Angle and Wind Momentum Ratios}\label{Canto}

\citet{Canto1996} present a convenient semi-analytic bow shock model solution for the shock resulting from the collision of two spherical winds at terminal speed.  In this model, the bow shock geometry and its surface density is determined by two fundamental ratios:

\begin{eqnarray}
\beta & = & \frac{\dot{M}_2v_2}{\dot{M}_1v_1},~{\rm and} \\
\alpha & = & v_2/v_1.
\end{eqnarray}

The distance of the bow shock from the star with the smallest momentum flux along the line joining the two stars, $R_0$ is given by 

\begin{eqnarray}
R_0={{\beta^{1/2}D}\over {1+\beta^{1/2}}}
\end{eqnarray}
and the asymptotic opening angle of the cone, $\theta _{\infty}$ by
\begin{eqnarray}
\theta _{\infty} - \tan \theta _{\infty}={\pi \over {1-\beta}}
\end{eqnarray}


\subsection{Continuum Linear Polarization of Colliding Wind Systems}

BME present a  complete theoretical construction for thin electron scattering in
generalized envelopes with an arbitrary
number of illuminating point sources.  The two main limitations of BME are in relation to finite star effects, such as occultation, and the explicit assumption of thin scattering.  For application to binary stars, the former can be an adequate assumption unless the binary is quite close or eclipsing.  Regarding the latter, the Thomson scattering cross-section is relatively small, and so thin scattering is often appropriate. However, for some high-density scattering regions, multiple scattering may modify the resulting polarization \citep{Hoffman2003,Shrestha2018}. We will employ both analytical models, described in detail below, and Monte Carlo simulations to investigate the full range of possible density distributions.

For massive star binaries, the high temperatures and strong ionizing
fluxes ensure the presence of copious free electrons.
Thomson scattering is a gray opacity, which suggests initially that
flat polarization is to be expected in the continuum.  
However, a flat polarization
is not generally true, particularly for a binary system.  

First, there can be changes across line features.
The influence of spectral lines come in 3 basic types:  (a) photospheric
lines, (b) recombination lines, and (c) scattering lines.  For
the scattering of starlight by electrons in a circumstellar envelope,
the polarization is flat across photospheric lines since the relative polarization is constant.  The line formation resides at the source, not in the extended circumstellar envelope.

For case (b), the recombination line is taken to form in the circumstellar
envelope.  It is possible for such line photons to be scattered by the circumstellar electrons.  But such photons are not emerging from the star
but distributed throughout the envelope, and represent a diffuse source
of photons for scattering.  Variations in polarization can be complex.
As an example, \citet{Ignace2000} explored variations in polarization at H$\alpha$
in a Be star disk with a 1-armed spiral density wave:  the polarization in the continuum is little affected by the antisymmetric structure, but in H$\alpha$ the density-squared emissivity leads to a highly non-symmetric radiation field
in the disk.  The result is distinct polarimetric behavior within the line as compared to the continuum.  Similar effects can occur for colliding wind systems \citep{Fullard2020thesis}.

There is an important limiting scenario for case (b) known as the ``line
effect'' \citep[e.g.,][]{1990ApJ...365L..19S}.  This is when the recombination line is formed at a large radius and the electron scattering occurs mainly over a more compact region close to the star.  In this situation the line photons are little scattered, and strong line emission can contribute to the flux $F_I$ but not $F_Q$ or $F_U$.  The result is a reduction in the relative polarization $p$ (sometimes referred to as ``diluted'' polarization).

Case (c) refers to the possibility of scattering resonance lines as an additional polarigenic opacity.  Although resonance scattering polarization for stellar winds has been explored in limited applications \citep{1989ApJS...71..951J,2000A&A...363.1106I}, the topic has lacked an observational driver.  Such effects require a high-resolution UV spectropolarimeter and the computational tools for making predictions of effects and providing fits to data.  The data has largely not existed, given that {\em WUPPE} was of lower spectral resolution.
{\em Polstar} will provide new possibilities for case (c), and the team possesses the radiative transfer tools for developing diagnostics that combine both electron and resonance line scattering.

More relevant for UV spectropolarimetry are the chromatic effects in the relative polarization that arise simply from the presence of two stars with different temperatures in a binary.  The net polarization from the unresolved binary systems involves a weighted linear sum of the polarized contributions from scattered starlight by each star.  Using 1 for the primary (more massive star) and 2 for the secondary (less massive star), the net polarization is of the form:

\begin{equation}
    p(\lambda) = f_1p_1+f_2p_2,
\end{equation}

\noindent where 

\begin{equation}
   f_1 = \frac{L_1(\lambda)}{L_1(\lambda)+L_2(\lambda)}   , 
\end{equation}

\noindent and

\begin{equation}
   f_2 = \frac{L_2(\lambda)}{L_1(\lambda)+L_2(\lambda)}    ,
\end{equation}

\noindent and $p_1$ and $p_2$ represent the relative polarization caused by geometric effects from the perspective of each separate star.  

Understanding $p_1$ and $p_2$ is aided with an example.  Suppose that the primary were surrounded by an axisymmetric disk.  This configuration would produce a polarization contribution $p_1$ if there were only one star.  However, that same disk is of course not centered on secondary, so starlight from the secondary scattered by the disk  produces a different polarization given by the parameter $p_2$.  For thin scattering the total polarization is just the weighted sum $p(\lambda)$.

Therefore, chromatic effects are predicted, despite Thomson scattering being gray.  If the stars have different temperatures\footnote{If the temperatures of the two binary components are identical, the polarized spectrum will be flat at every wavelength, in the absence of radiative transfer effects.}, their spectral energy distributions (SEDs) will differ.  This leads to a polarized spectrum, $p(\lambda)$, that in general is not flat.  In particular, for hot massive stars, the deviation from a flat polarized spectrum occurs mainly in the UV.  Thinking simplistically in terms of Planckian stellar spectra, hot OB star spectra in the optical and at longer wavelengths is in the Rayleigh-Jeans regime\footnote{The situation is different for WR star that do not follow the Rayleigh-Jeans law owing to significant
 free-free and bound-free emission from the wind.}.  In those wavebands, $f_1$ and $f_2$ become constants.  But in the UV, where {\em Polstar} will operate, the presence of differing Wien peaks ensure that $f_1$ and $f_2$ vary with wavelength, and $p(\lambda)$ will not be flat.  Formally, $p(\lambda)$ is bounded by $p_1$ and $p_2$, which provides a richer set of constraints for inferring the geometry of the system. Specifically in terms of colliding winds, the UV in general provides new perspectives with spectropolarimetric data relating to the relative location and geometry (e.g., opening angle) of the colliding wind interaction (CWI) region.

For our application to massive colliding wind binary systems, we assume the colliding wind interaction (CWI) is axisymmetric about the line-of-centers joining the two stars.  We further assume that the separate winds of the two stars are each spherically symmetric up to the CWI.  As a result, in the notation of BME, we have $\gamma_1=\gamma_2=\gamma_3=\gamma_4=0$,
and the only factors that are nonzero are $\gamma_0$ and $\tau_0$.  

BME then considered the more
limited scenario of a binary system with a circular orbit and corotating
envelope.  Our approach allows for elliptical orbits (which \citealt{Brown1982} later
considered), and using Section \ref{Canto}, we
adopt an analytical solution for the CWI from \citet{Canto1996}.

Our team possesses a suite of computational tools for time-dependent hydrodynamical simulation of colliding winds \citep[e.g.,][]{2017MNRAS.464.4958R} and for the radiation transport for synthetic polarization spectra \citep{Hoffman2007,Huk2017}.  Here we explore the BME approach for several reasons.  First it generally has applicability to OB+OB star winds where the electron scattering does tend to be thin.  It can also have applicability to WR+OB when the CWI is relatively far removed from the WR star so that electron scattering is again optically thin.  Finally, solutions for CWIs are complex, depending on a number of variables pertaining both to the stars and to the orbit.  BME provides convenient analytic solutions to allow for exploration of a large parameter space.  
However, for the sake of brevity, we do assume 
axisymmetry about the line of centers joining the two stars.  This is not always the case owing to the Coriolis force \citep{2007ApJ...662..582L}.  Also our notation departs slightly from BME, although we still employ similar variables.  

\subsection{Model Prescription}

In our model, the winds of the two stars and the intervening colliding
wind shock are prescribed using primarily polar coordinates for
each star.  The primary star\footnote{By ``primary star'', we refer to the star with the stronger wind momentum, meaning that when the winds are unequal, the bowshock envelopes the secondary star.} is assumed to have the stronger wind
in terms of momentum flux, $\dot{M}v_\infty$, where $\dot{M}$
is the mass-loss rate and $v_\infty$ is the terminal wind speed.
The secondary is then the weaker wind case in terms of this
product.  We typically use subscripts ``1'' and ``2'' for
the primary and secondary.

We define the polar coordinates with respect to the axis that is the line
of centers between the two stars.  Therefore the polar coordinates centered on the primary star 
$(r_1,\theta_1,\phi_1)$ are such that $\theta_1 = 0$ in the direction
of the secondary.  Likewise, the coordinates linked to the secondary star 
$(r_2,\theta_2,\phi_2)$ also have $\theta_2 = 0$ in the direction
of the primary.  Frequently, our approach uses the standard cosine
notation, $\mu_1 = \cos \theta_1$ and $\mu_2 = \cos \theta_2$.

\begin{figure}
\plotone{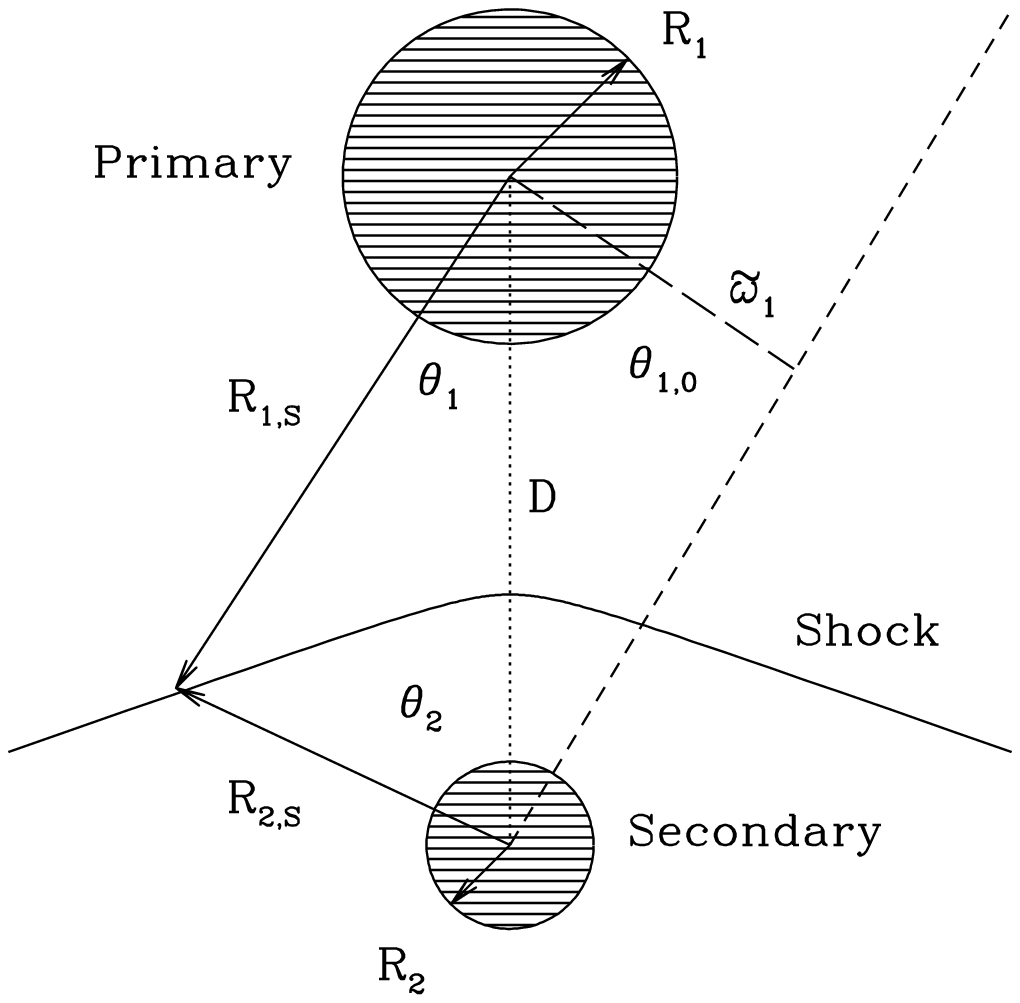}
\caption{Top-down illustration of the two stars (primary and secondary) and
the bowshock formed by the colliding winds.  Labeled variables are defined
in text.
\label{fig1}}
\end{figure}

The individual winds are taken
to be spherical with densities varying with the inverse square of the distance from the star.  There is a vast range of binary systems.  In some the CWI forms at radii where both winds are at terminal speed.  In others one wind is at terminal speed, and the bow shock forms in the wind acceleration of the other star (perhaps even crashing onto the hydrostatic atmosphere).  To explore polarimetric variability from colliding wind shocks, we
use the convenient formulation of \citet{Canto1996} for their semi-analytic bow shock
solution for two spherical winds at terminal speed (see Section \ref{Canto}). 

\noindent Figure~\ref{fig1} provides a schematic of the binary
system with intervening CWI region between the two stars.
The radial distance of the bow shock from either star
is denoted as $R_{1,S}$ and $R_{2,S}$ given by 

\begin{eqnarray}
R_{2,S} = D\,\frac{\sin \theta_1}{\sin(\theta_1+\theta_2)},~{\rm and} \\
R_{1,S} = \sqrt{D^2+R^2_{2,S}-2\,D\,R_{2,S}\,\cos\theta_1},
\end{eqnarray}

\noindent where $D$ is the separation between the two stars at any moment.
Naturally the solution is only valid when $D>R_{2,\ast}$.
A special quantity is the standoff radius of the bowshock along
the line centers, denoted as $R_{1,0}$ and $R_{2,0}$.

\begin{eqnarray}
R_{2,0} = \frac{\beta^{1/2}}{1+\beta^{1/2}}\,D,~{\rm and} \\
R_{1,0} = \frac{1}{1+\beta^{1/2}}\,D.
\end{eqnarray}

\noindent The case $\beta=1$ corresponds to a planar shock between
identical stars and winds, with $\theta_{1,\infty}=\theta_{2,\infty}=\pi/2$.

For a particular geometry, as expressed by the
wind and orbital properties, $p_1$ and $p_2$ have generally different values
but are defined with respect to the same line of centers joining the two stars.  
While these parameters are not chromatic, as noted above the total polarization
$p_{\rm tot}$ can
be chromatic because the two illuminating sources will generally have
different SEDs, with  

\begin{equation}
p_{\rm tot} = \frac{L_1(\lambda)p_1+L_2(\lambda)p_2}{L_1(\lambda)+L_2(\lambda)}.
\end{equation}

\noindent This result, in our notation, is equivalent to equations~(6a)
and appropriate lines of (7) from BME for our axisymmetric geometry.
\\
\\
\\

\subsection{Model Results}

\subsubsection{Chromatic Effects in the UV}

To illustrate chromatic effects, we fixed a particular set of binary parameters
while allowing the temperatures of the two stars to vary.  The fixed properties lead to $p_1=0.009$ and $p_2 = 0.019$.
In this example, the binary components are separated by $40~{\rm R}_\odot$.  The secondary
is twice as large as the primary; the two winds have equal speed; the mass-loss
rate for the primary is $3.3\times$ larger than the secondary.  Additionally,
we took $\sin i = 1$.

\begin{figure}
\plotone{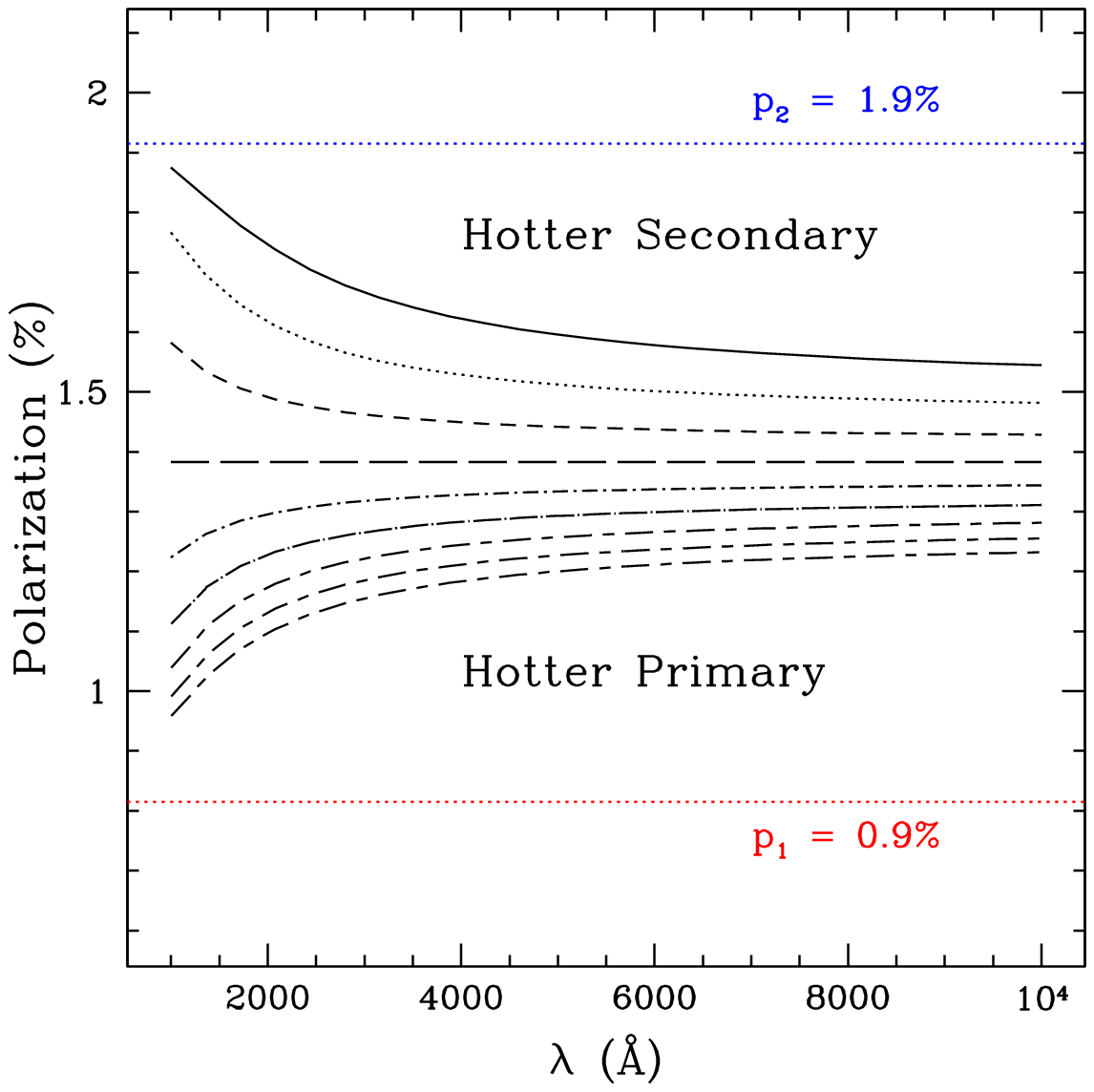}
\caption{Variation of the polarized continuum with wavelength, here shown from the FUV
to 1~micron.  The stars are treated as Planckian.  The temperature of the secondary is
fixed at 25,000~K, and the temperature of the primary varies from 16,000~K to 40,000~K
in 3,000~K increments.  The particulars of the stellar and wind parameters for this
illustrative case are described in the text.  For the selected parameters, the limiting polarizations $p_1$ and $p_2$ are indicated with horizontal red and blue dotted lines,
respectively.
\label{fig5}}
\end{figure}

To obtain the spectropolarimetric continuum shape, we treated the two stars as  Planckian sources with effective temperatures $T_1$ and $T_2$.
We fixed the temperature of the secondary at $T_2=25,000$~K.  We varied the temperature
of the primary from $16,000$~K to $40,000$~K in $3,000$~K
intervals, and display the results in Figure~\ref{fig5}, where polarization is
shown as positive.  For purposes of illustration,
the usage of ``primary'' and ``secondary'' becomes ambiguous as one of the temperatures
varies but all other properties remain the same.  The point is to highlight
the fact that when the stars have different spectral energy distributions, the
continuum polarization is not generally constant with wavelength, even though electron scattering
is gray.  Only when the two stars have equal temperatures is the continuum polarization
truly flat.

Note especially that the wavelength varies from the FUV through the optical to 1 micron.
For massive stars with typical temperatures well in excess of $10,000$~K,
both the primary and secondary can be characterized as following the Rayleigh-Jeans law in the optical, and  the continuum is always flat.  It is only in the
UV that the polarization deviates significantly from a flat profile.  For the selected
parameters, the polarization actually drops toward the UV when the primary
is hotter (i.e., more luminous), owing to the fact that $p_1 < p_2$.
By contrast, when the secondary is hotter (i.e., more luminous), the polarization
increases significantly.  Ultimately, for any combination of binary parameters,
when the more luminous star in the UV also has the higher polarimetric component, the polarization is enhanced in the UV, and vice versa.

\subsubsection{Variable Polarization with Orbital Phase}

The polarimetric properties of the colliding wind system depend on the binary separation, $D$.  For a circular orbit, the values $p_1$ and $p_2$ are constant.  This is not true when the orbit is eccentric.  For eccentricity $e$ and semi-major axis $a$, the binary separation varies as

\begin{equation}
D(\varphi) = a\,\frac{1-e^2}{1+e\cos\varphi},
\end{equation}

\noindent where $\varphi$ is the orbital azimuth, defined so that
$\varphi=0$ corresponds to periastron.  Thus $p_1$ and $p_2$ are
functions of orbital phase through the variation of $D(\varphi)$, since even for fixed $\alpha$ and $\beta$ ratios, the surface density of the CWI changes as well as the extent of the wind for each respective star up to either side of the CWI.

To obtain the variable polarization with orbital phase, we define $i_0$ as the viewing inclination of the orbital plane, so that $i_0=0^\circ$ is a top-down view and $i_0=90^\circ$ is edge-on. 
The polarimetric variability is determined by

\begin{eqnarray}
q & = & p_{\rm tot}(i)\,\cos 2\psi,~{\rm and} \\
u & = & p_{\rm tot}(i)\,\sin 2\psi,
\end{eqnarray}

\noindent where the polarization position angle $\psi$ relates to orbital azimuth and orbital plane inclination as

\begin{equation}
\tan \psi = -\cos i_0/\tan \varphi.
\end{equation}

\noindent The inclination of the line of centers to the viewer's line of sight is given by

\begin{equation}
\cos i = \sin i_0\,\cos \varphi.
\end{equation}

\begin{figure}
\centering
\includegraphics[width=0.9\columnwidth]{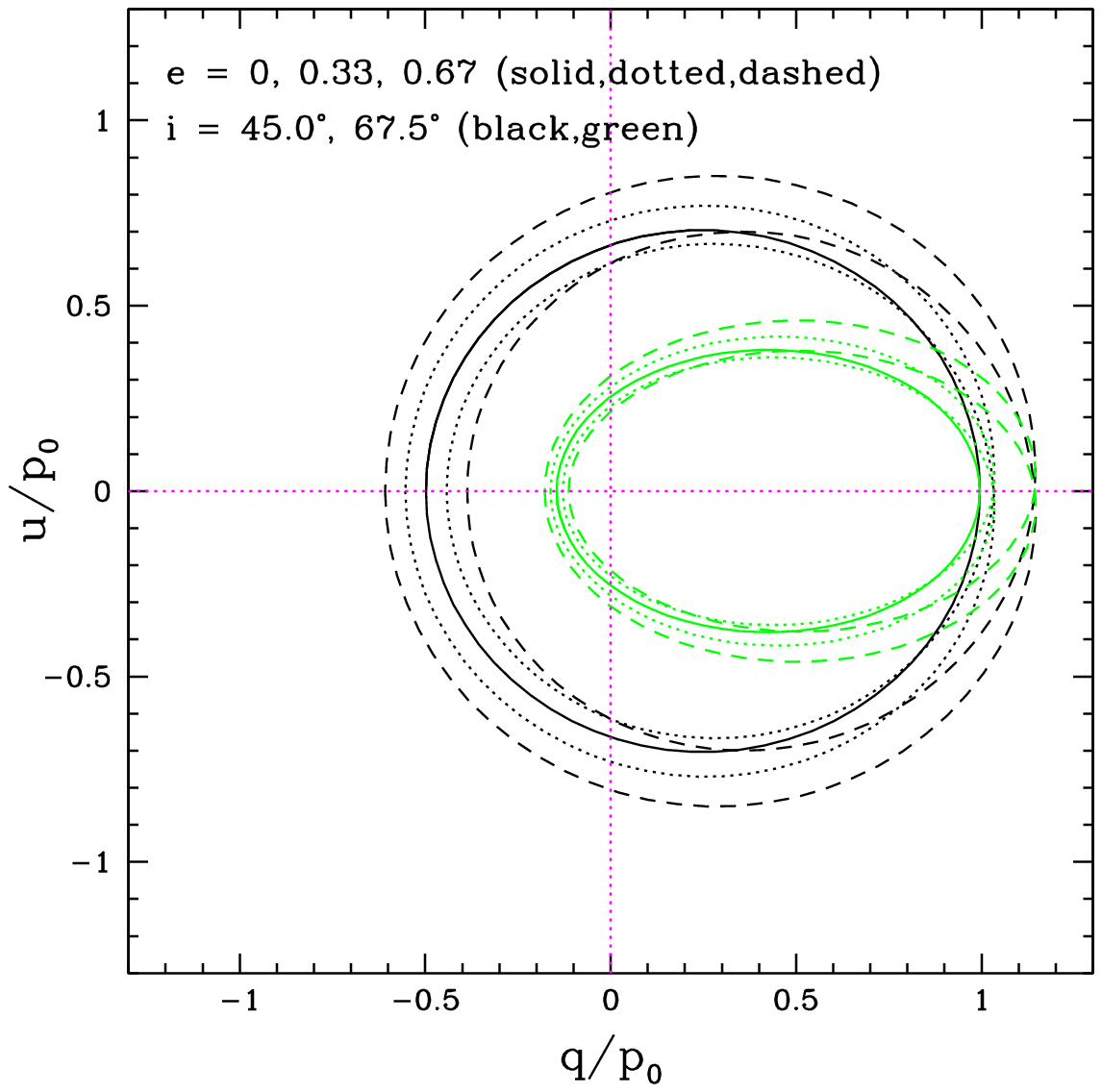}
\includegraphics[width=0.9\columnwidth]{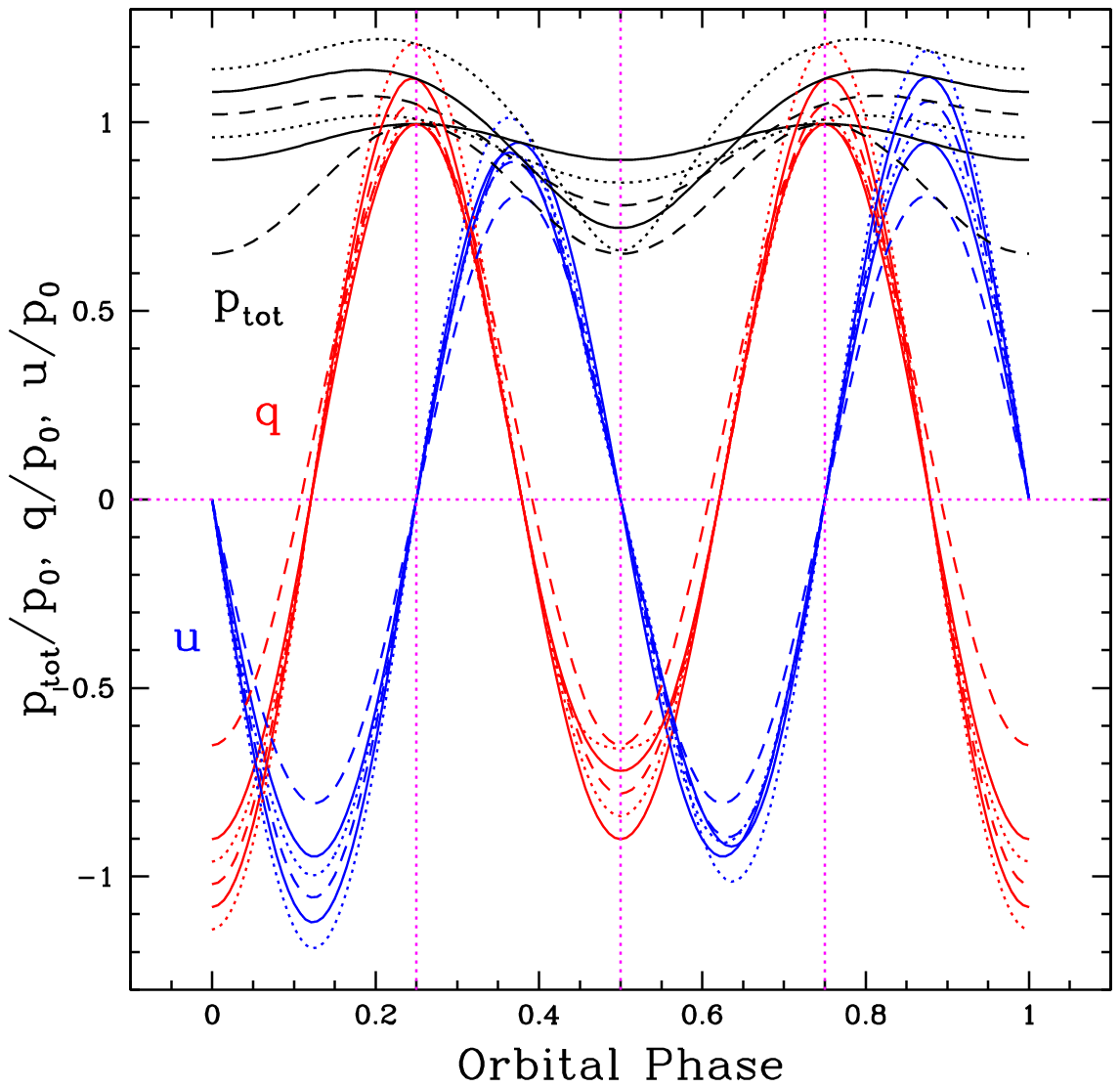}
\caption{(Top) Example $q-u$ loops for a binary consisting of identical stars, for which $p_1=p_2\equiv p_0$.  Eccentricities $e$ and inclinations $i$ are indicated for line type and color.  There are two loops per orbit, but with eccentric orbits, the pair separate into different sizes except for $e=0$.  (Bottom) Same 6 models as top, now displayed as light curves in $q$ (red), $u$ (blue), and total polarization $p_{\rm tot}$ (black).  Periastron passage is at phase 0.0, and apastron at phase 0.5}
\label{orbitpol}
\end{figure}

Figure~\ref{orbitpol} displays a suite of polarimetric variations for binary systems with different values of $e$ and $i_0$. We fitted a linear regression to $p_{\rm tot}(D)$ in the case of a binary with two identical stars (i.e., planar CWI shock) to obtain $p_0 = 1.328 - 0.0111D$, normalized so that $p_{\rm tot} = 1$ at $D=30~R_\odot$.  We show model variable polarization curves for  inclinations 
and  eccentricities, as labeled.  Top is for $q-u$ loops; bottom shows polarized light curves with orbital phase. 
The line types represent the different eccentricities. 
At $i=90^\circ$ (not shown), all curves would become horizontal lines with only $q$ variation but no $u$ variation.  Note also that $p_{\rm tot}=\sqrt{q^2+u^2}$.

\vskip 1truecm

\subsubsection{WR+O Binaries}

\begin{figure}
\plotone{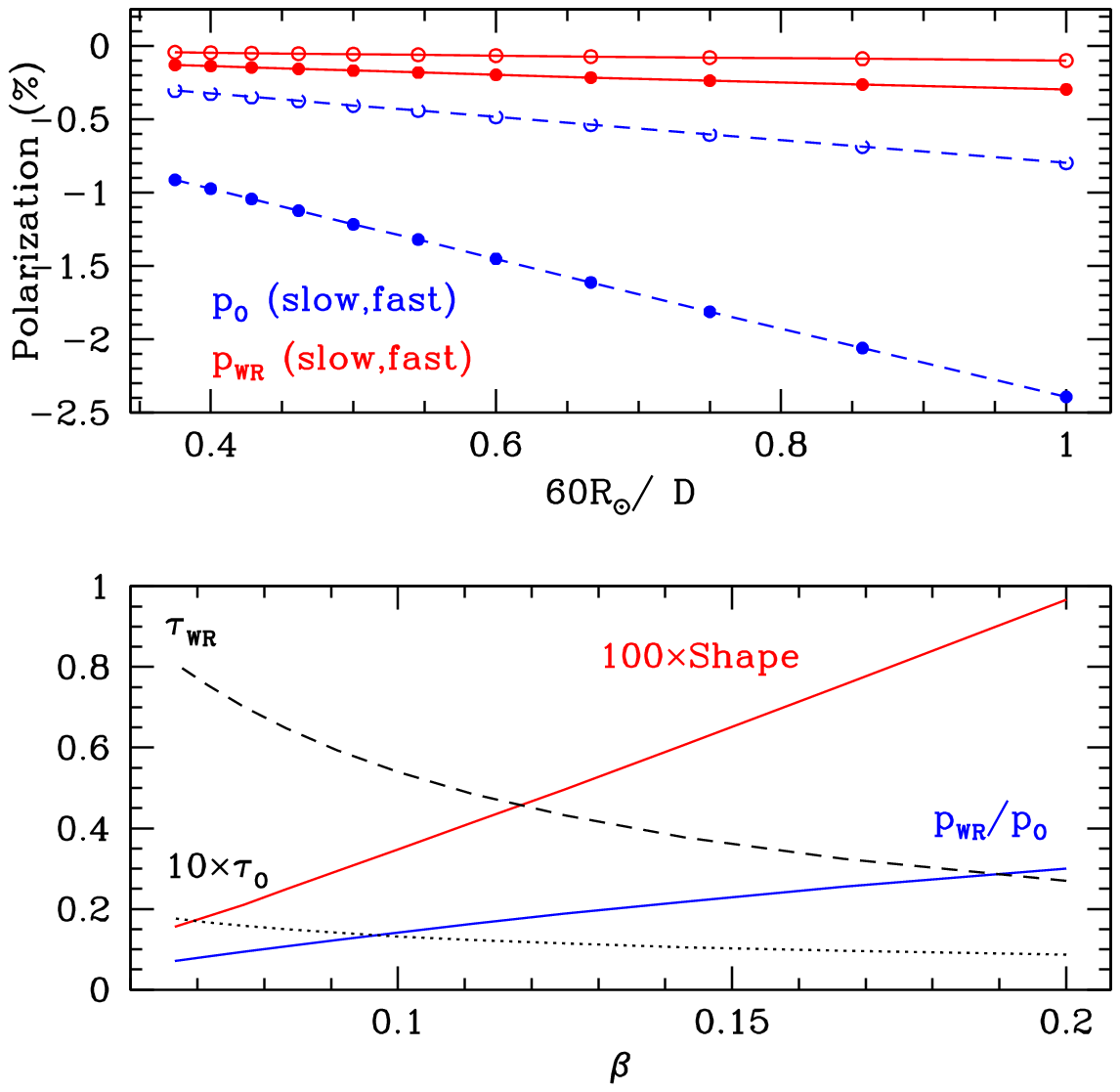}
\caption{Model results for a WR+O star binary (Section 3.4.3).  
The upper panel shows results for a fixed value of $\beta=0.1$ for the slow and fast wind
cases, plotted against binary separation as $D^{-1}$.  Blue lines represent $p_O$ and red
represent $p_{WR}$.  The lower panel shows results for $D=160~R_\odot$ with $\beta$ varying
between 1/15 and 1/5.  Black curves represent the WR and O optical depths, as labeled.  
The blue line in this panel represents the ratio of $p_{WR}/p_O$.  The red line represents the ``shape,'' defined as the ratio
$(1-3\gamma_O)/(1-\gamma_{WR})$ and scaled up by $100\times$.
\label{fig8}}
\end{figure}

We calculated $p_{WR}$ and $p_O$ for three scenarios, with a summary of results displayed in Figure~\ref{fig8}.  The first case is $\beta=0.1$ with orbital separation ranging from $60~R_\odot$ to $160~R_\odot$ for ``slow'' winds of 1000 km s$^{-1}$ for both stars.  The second case is for fast winds at 3000 km s$^{-1}$, with all other parameters fixed.  The third scenario corresponds to an intermediate wind speed of 2000 km s$^{-1}$ at a fixed separation of $D=160~R_\odot$, but with $\beta$ ranging between 1/15 and 1/5.

The upper panel of Figure~\ref{fig8} summarizes the comparison between slow and fast
winds.  Note that polarizations are negative for our convention of a binary with orbital plane seen face-on, with the line of centers oriented north-south in the sky.  In this panel the red lines represent $p_O$ and blue represent $p_{WR}$.  Solid lines are for the fast wind case, and dashed for the slow wind case.  The results are plotted against $D^{-1}$, normalized as indicated.  The net result is that the polarization is overall larger for a slower wind, since the density is larger.  We find that $\tau_{WR}$ is roughly constant as $D$ changes, indicating that its value is dominated by the relatively extended spherical wind of the WR~star, since the CWI is far removed. Since the CWI is relatively farther from the WR~star with increasing $D$, $(1-3\gamma_{WR})$ becomes smaller with $D$.  Consequently, $p_{WR}$ decreases with increasing $D$.
The behavior for the O~star is that the polarization is dominated by the CWI.  The surface density of the CWI shock for the \citet{Canto1996} solution scales as $D^{-1}$ overall.  These is
evidenced by the fact that both blue curves appear quite linear in the plot.

For the lower panel of Figure~\ref{fig8}, we display the results differently as $\beta$ is allowed to vary between 1/15 and 1/5 and with smaller $\beta$, the CWI is closer to the O~star component.  Consider first the dashed and dotted curves in black, for $\tau_{WR}$ and $\tau_O$, respectively.  As $\beta$ becomes smaller, $\tau_{WR}$ is larger, approaching the limit of the strictly spherical wind value.  The value of $\tau_O$ is much lower, and is plotted as scaled up by $10\times$.

The blue curve in this lower panel represents the ratio of $p_{WR}/p_O$. Its behavior indicates that from geometrical considerations, the contribution to the polarization from the O~star wind is much greater than for the WR~wind, even more so as $\beta$ becomes smaller.  Even though the WR~wind has a much higher optical depth scale, the distortion of the scattering envelope from spherical is quite minor  from the perspective of the WR~star.  This is made clear by the red curve, where ``shape'' is the ratio $(1-3\gamma_{WR})/(1-3\gamma_O)$, and scaled up by $100\times$.  From the perspective of the O~component, the scattering envelope is highly distorted.  

In combination, these results suggest that at wavelengths where the O~star is more luminous, the polarization will overall be larger (biased toward $p_O$) than at wavelengths where the WR~star is more luminous (polarization biased toward $p_{WR}$).  Our treatment does have deficiencies, the most important being that we ignore the wind acceleration region, and that we treat the WR~wind as optically thin to electron scattering.  For the latter, it is clear that the WR~wind is already a minor contributor to the polarization when $\beta \ll 1$, and a more full treatment of multiple scattering is hardly expected to impact that conclusion.  For the former, inclusion of the wind acceleration region and associated density distribution, along with radiative inhibition could certainly change the detailed outcomes.  Nonetheless, the present treatment indicates that $p_O \gg p_{WR}$, another qualitative result that is unlikely to change despite our more simplistic assumptions.

\subsection{Linear Polarization Across Spectral Lines}

At optical and UV wavelengths it is often a good approximation to assume that electron scattering is coherent in the frame of the electron. However, since the electron is moving (due to both thermal and large scale motions) relative to both the source and observer there will be a wavelength shift in the
observers frame \citep{Auer1972}. Thermal motions can lead to either a decrease or increase in the scattered photon's wavelength, while scattering by an expanding  monotonic flow will always lead to a redshift.
These velocity shifts are observed  -- in P~Cygni for example, in which the thermal electron velocities are
larger than the wind velocity, the Balmer series, for example, show nearly
symmetric broad extended wings centered on the emission profile \citep{Bernat1978}. By contrast, in the spectra of many WR stars only a red electron scattering wing is seen, since in these stars the wind velocities are larger than the electron thermal velocities \citep{Hillier1984}.

\begin{figure*}[t]
\centering
\includegraphics[width=12 cm,angle=90]{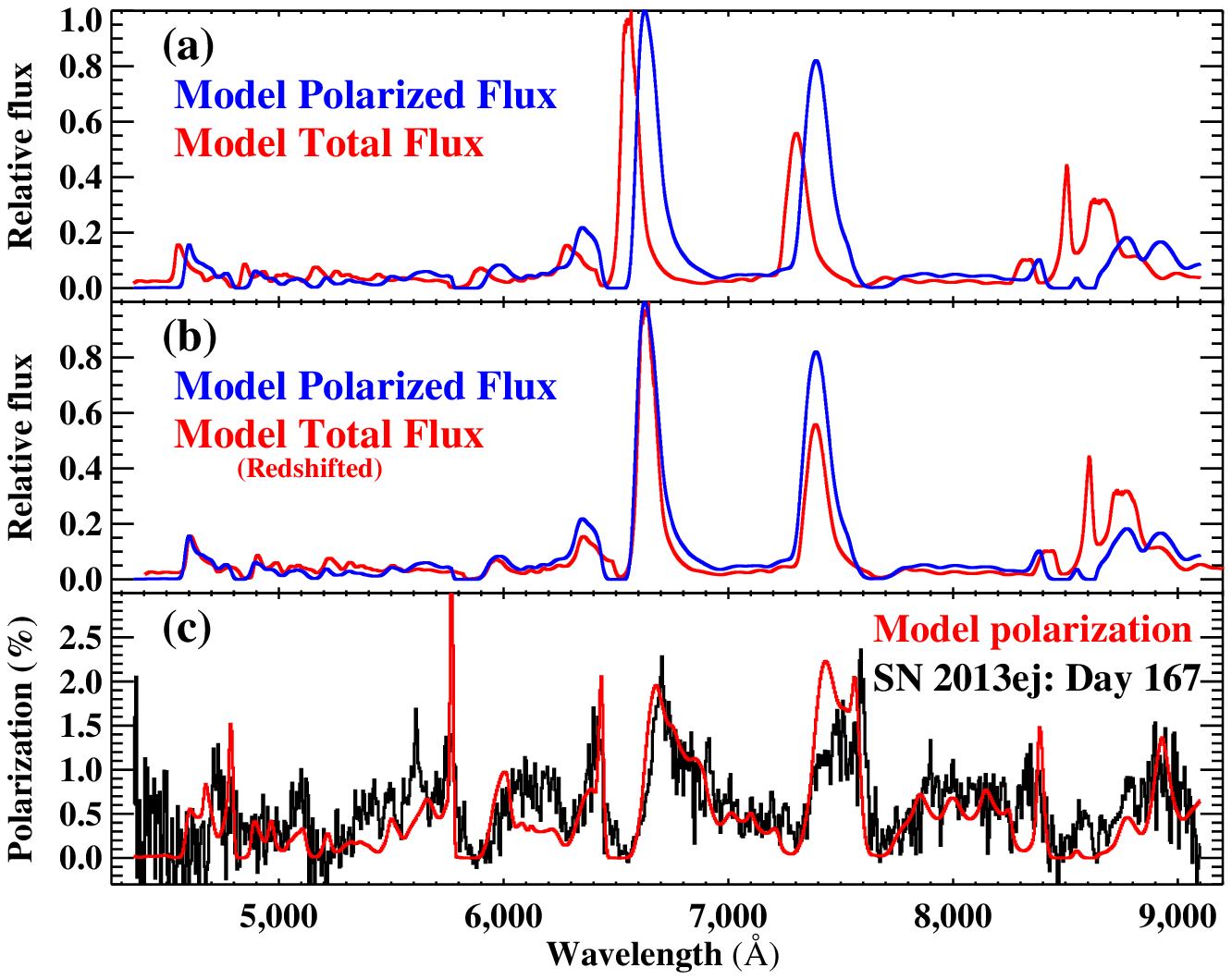}
\caption{Illustration of how velocity shifts influence the polarized spectrum. Top panel shows how the scattered spectrum (blue) is offset to the red from the flux spectrum (red). The second panel corrects for the offset, illustrating the similarities between the two spectra. However, because spectral features arise over a range of radii, the agreement is not perfect. The bottom panel shows the good agreement between a model (red) and observation (black) for SN 2013ej (Figure from \cite{Leonard2021}; reproduced by permission of the AAS).}
\label{Fig_SN_pol}
\end{figure*}

For continuum polarization the velocity shifts induced by electron scattering are generally unimportant. However, they are of crucial importance in understanding line polarization. In Fig.~\ref{Fig_SN_pol} we show an extreme
example, taken from polarization studies of supernovae, where the line shift is crucial. In the Type II 
SN 2013ej the polarized spectrum is very similar to the observed spectrum \citep{Leonard2021}. However, it is different in two important ways: First,  the spectrum is redshifted. Second, the agreement between the intrinsic and polarized spectrum depends on which spectral feature is being examined.   Both effects are easily understood by assuming that the scattering source is offset from the SN emitting region, and by noting that
line emission in a SN region is stratified (different lines originate in different regions of the SN ejecta).  \cite{Dessart2021} argued that scattering region was due to a nickel bubble, with an enhanced electron density
offset from the center of the ejecta. Due to motion relative to the source, the
spectrum is shifted to the red. Furthermore, the similarity between the intrinsic and scattered light will generally be 
highest for lines arising in the interior regions of the SN ejecta, since in that case the scattering angle is the same, unlike the case where the emission is formed over a much larger volume.

When optically thin electron-scattering is assumed, and other absorption processes neglected (e.g., purely bound-bound transitions), 
models predict that for an oblate spheroid the polarization should be parallel to the major axis. The reason is simple -- more photons will be scattered from the equatorial regions (with their
electron vector parallel to the major axis) than the polar regions. However, when optical depth effects are allowed for, the polarization can flip sign, as scattered photons become 
more isotropic in the higher optical depth equatorial region.

\begin{figure}
\includegraphics[width=8 cm]{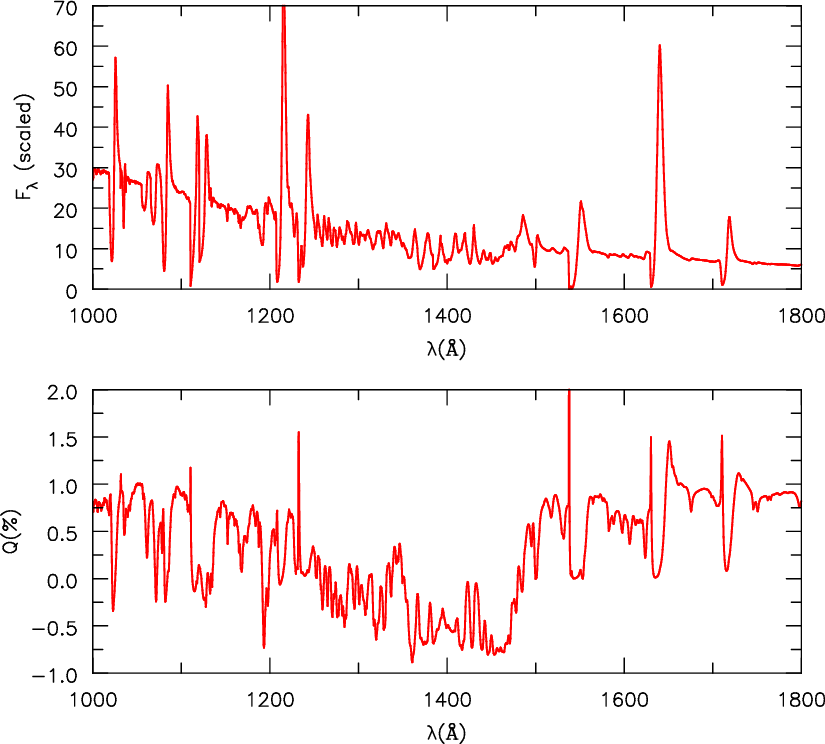}
\caption{The spectrum and polarization (Stokes Q) for a WR star with an oblate stellar wind viewed edge on. The ratio of the polar to equatorial density is 0.3. Notice how the polarization switches sign -- a consequence of optical depth effects in the wind.}
\label{Fig_WN_pol}
\end{figure}

The above scenario is illustrated in \ref{Fig_WN_pol} where we illustrate the spectrum for a WN Wolf-Rayet star assuming it has an oblate wind with a polar to equatorial density ratio of 0.3. Near 1800\,\AA, the polarization is positive, as expected for an oblate spheroid. However, around 1400\,\AA, the polarization is negative, a consequence
of Fe-blanketing, which increases total opacity and also introduces line effects.
The assumed axial symmetry guarantees that the position angle must be parallel or perpendicular to the rotation axis, so here we have taken the 
Stokes $Q$ component to subtract light polarized perpendicular
to the rotation axis from light polarized along the axis (and the $U$ component 
at 45 degrees to that would be zero in this symmetry).

Both of the above examples illustrate how the polarization contains important information about the structure of
circumstellar matter, whether it be a SN ejecta or a wind. Importantly, we now have the numerical tools to
interpret these observations. Polarization, in conjunction with other observations provides an invaluable tool to
help understand astrophysical objects, especially since most are spatially unresolved with modern telescopes.

\subsubsection{Numerical models}

Radiative transfer models are capable of modeling more complex binary phenomena than those allowed by the assumptions of BME \citep[e.g., ][]{Kurosawa2002,Hoffman2003}. Leveraging these models is essential for any deep investigation of specific colliding wind systems. Polarization of spectral lines is especially suited to numerical modeling because of how difficult the problem is in asymmetric geometries. 

\begin{figure}
\centering
\includegraphics[width=0.9\columnwidth]{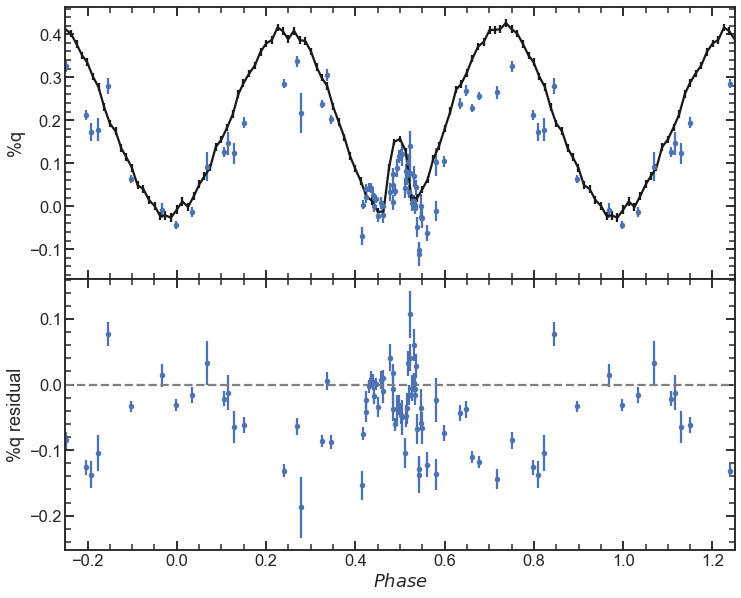}
\includegraphics[width=0.9\columnwidth]{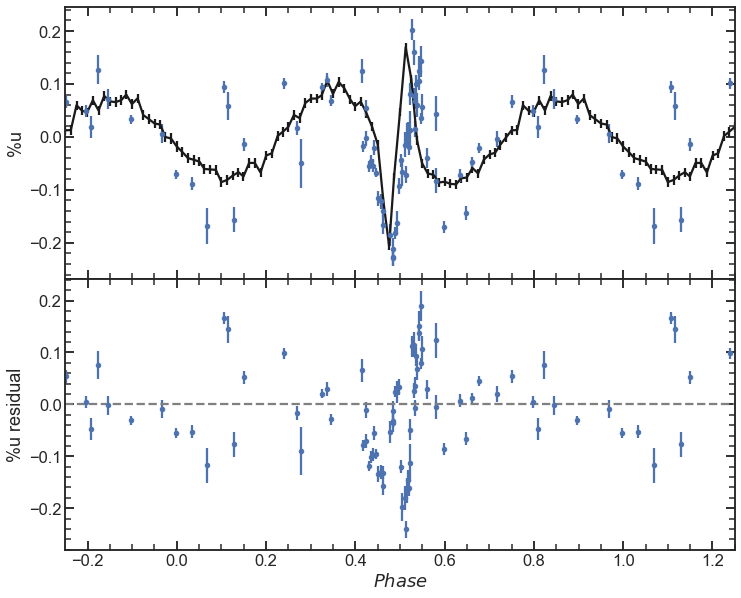}
\caption{V444 Cygni polarization (blue points) compared to the \citet{Kurosawa2002} E model at an inclination angle of 82.2$^{\circ}$ (black line), with  rotated O star wind cavity as in the emission line model. The $R$-band polarization data are taken from \citet{St.-Louis1993}. Residuals are also presented below each plot.}
\label{fig:v444_toy_continuum}
\end{figure}

Numerical models can also be used as initial probes much like BME. Because electron scattering is the primary polarigenic mechanism in hot star winds, the effects of line and continuum polarization can be decoupled by considering their emission and scattering regions. A toy model of the well-studied V444 Cyg has been constructed for this purpose \citep{Fullard2020thesis}. It is based on the \citet[][]{Kurosawa2002} E model of the system including a WR star with an appropriate wind density profile from the Potsdam Wolf-Rayet (PoWR) models\footnote{See the Potsdam PoWR models at the following website: www.astro.physik.uni-potsdam.de/~wrh/PoWR/powrgrid1.php}. A simple spherical cap cutout represents the O star wind (assumed to be of negligible relative density) at the wind collision point, rotated to mimic the effect of orbital motion. Emission occurs from the WR and O star photospheres (where the WR star photosphere is defined from its PoWR model). It reproduces the continuum behaviour of the system as shown in Figure~\ref{fig:v444_toy_continuum}. 
 
\begin{figure}
\centering
\includegraphics[width=\columnwidth]{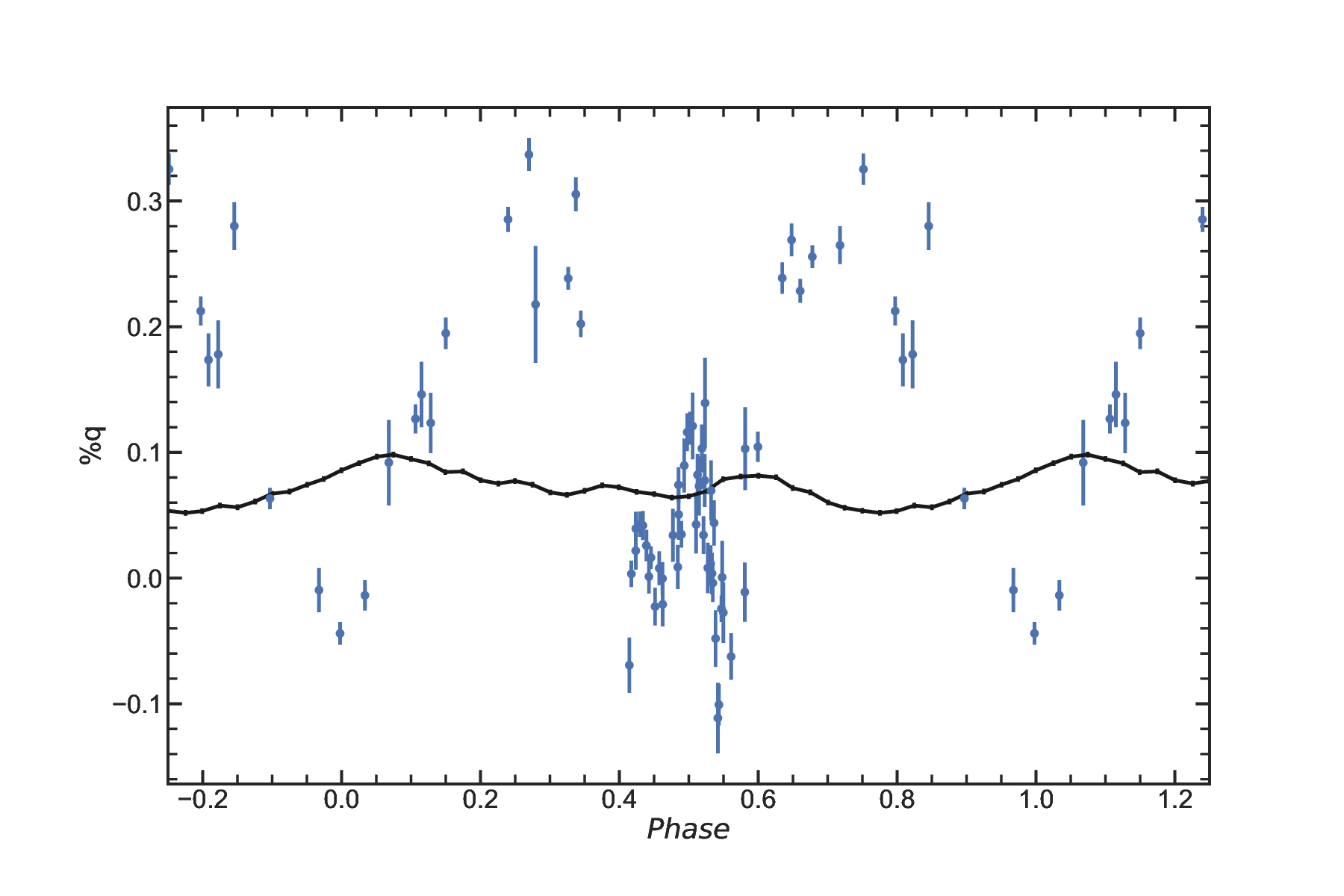}
\includegraphics[width=\columnwidth]{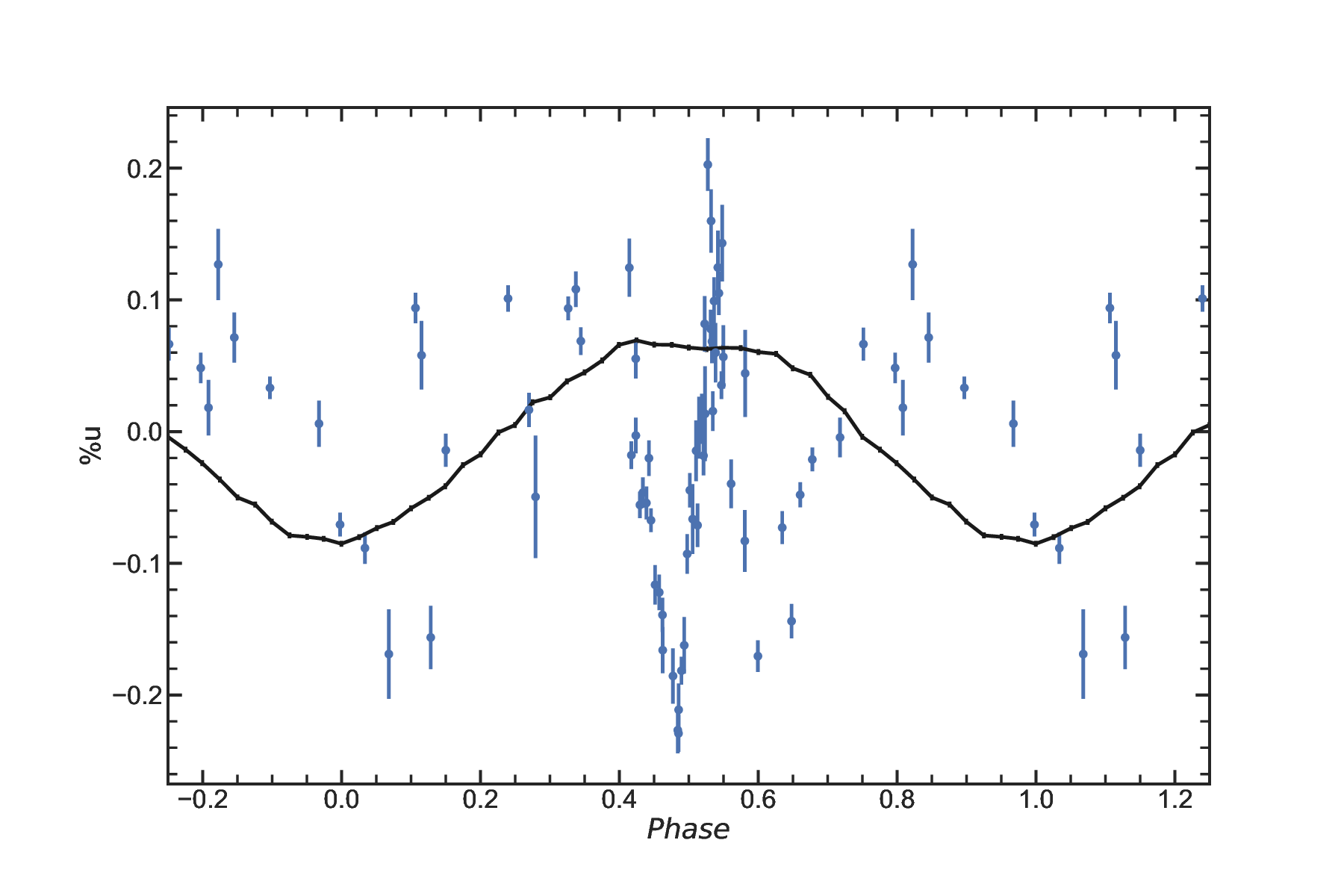}
\caption{V444 Cygni polarization (blue points) compared to a wind + shock emission line region model at an inclination angle of 82.2$^{\circ}$ (black line). The $R$-band polarization data are taken from \textcite{St.-Louis1993}. The comparison between the expected signal in a He{\sc ii} line from a wind and CWI and the observations in continuum light clear shows that the signatures are completely different but are both of similar amplitude (in this case only in u).}
\label{fig:stlouis_windshock_slip_compare}
\end{figure}

\begin{figure}
\centering
\includegraphics[width=\columnwidth]{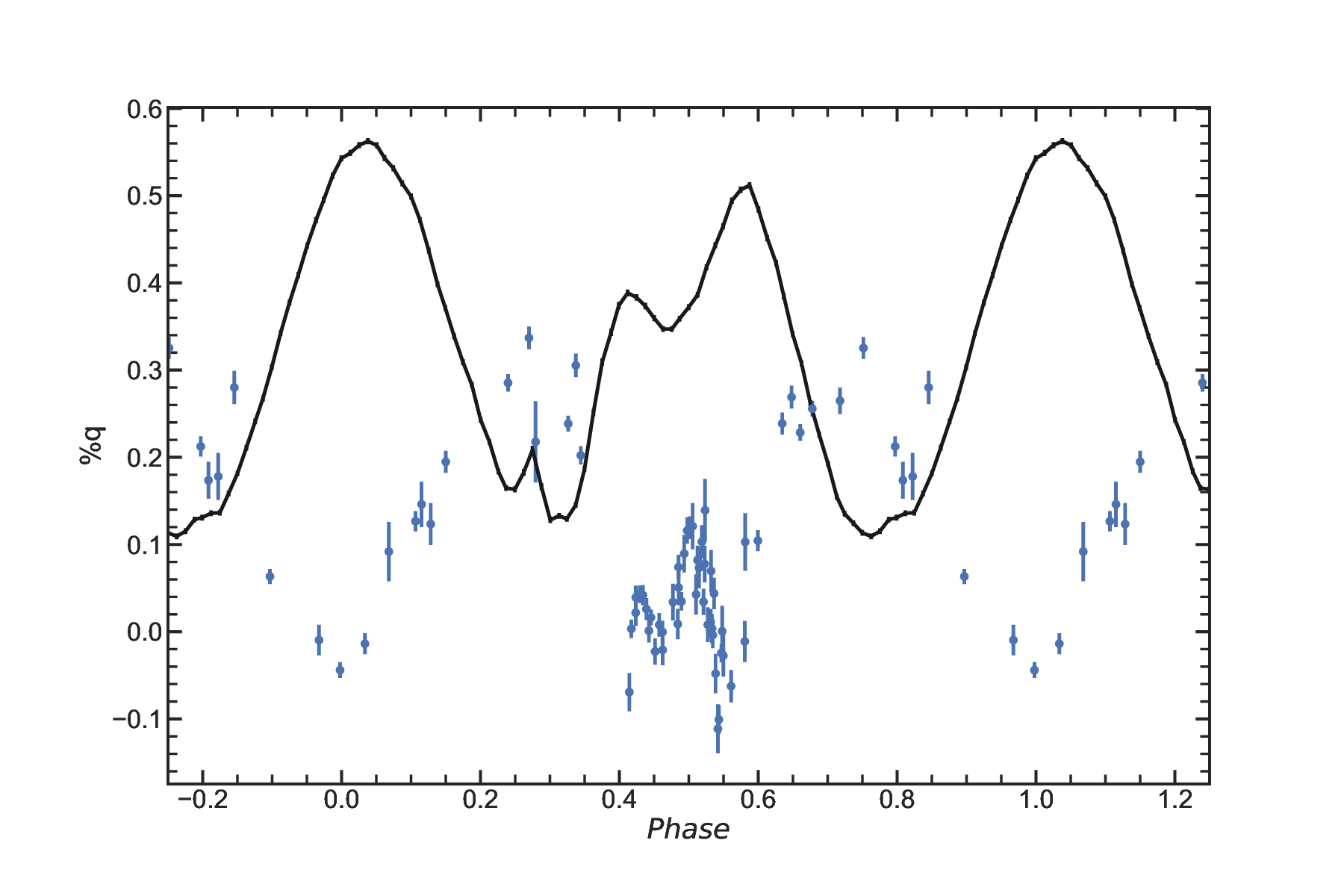}
\includegraphics[width=\columnwidth]{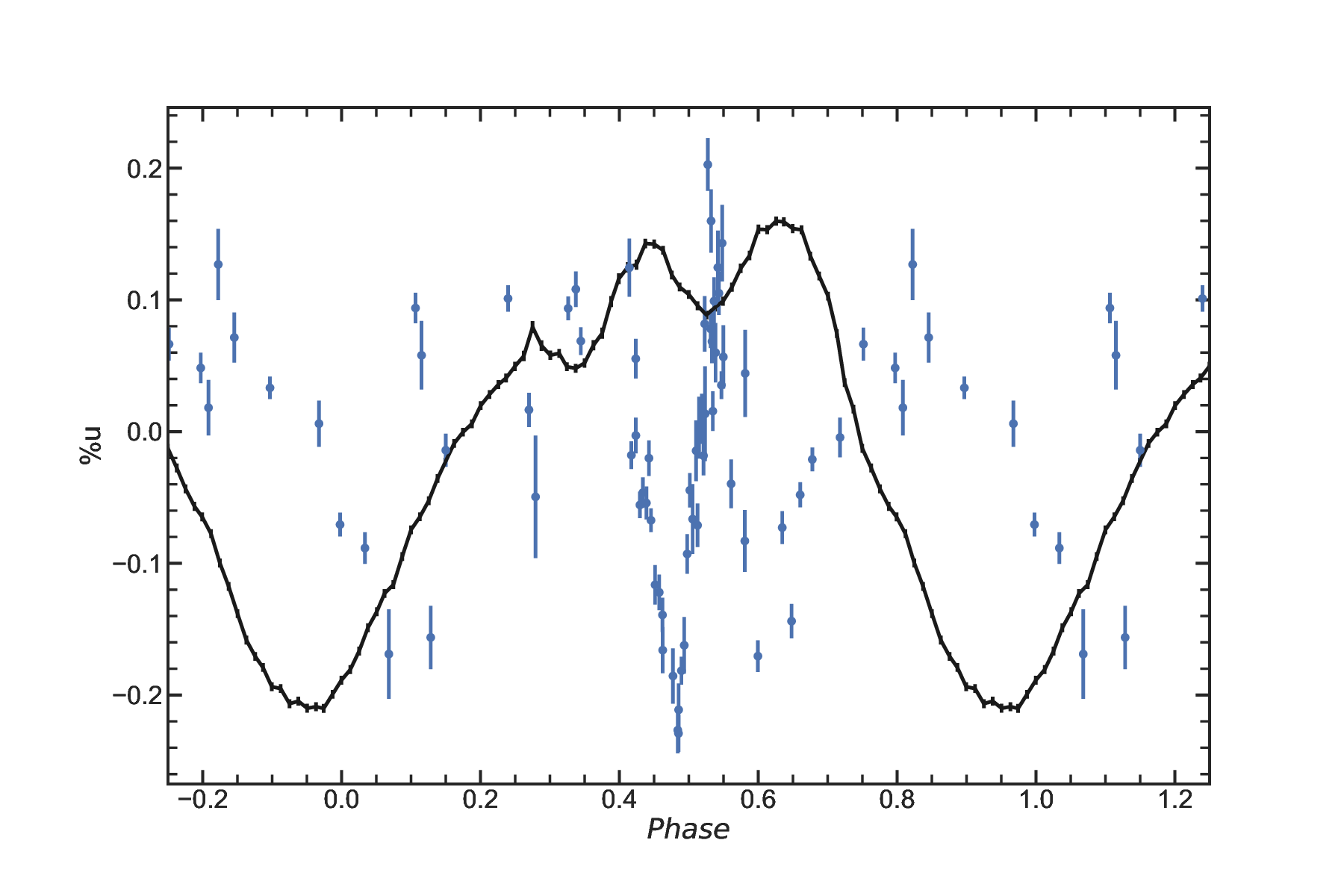}
\caption{V444 Cygni polarization (blue points) compared to a shock emission line region model at an inclination angle of 82.2$^{\circ}$ (black line). The $R$-band polarization data are taken from \textcite{St.-Louis1993}.The comparison between the expected signal in a He{\sc ii} line from a CWI and the observations in continuum light clear shows that the signatures are completely different but are both of similar amplitude}
\label{fig:stlouis_shock_slip_compare}
\end{figure}

To represent line polarization caused by He{\sc ii}, the emission locations are simply changed to originate from a shell in the WR star wind, and from the edge of the shock region in two clumps. The clumps are located at phase $\sim0.12$ and $\sim0.75$, and lie at the edge of the simulation just beyond the O star orbit. Figures~\ref{fig:stlouis_windshock_slip_compare} and~\ref{fig:stlouis_shock_slip_compare} show the resulting polarization curves compared to existing continuum polarization data. Figure~\ref{fig:stlouis_windshock_slip_compare} shows the complete wind and shock model, which has a low near-constant Stokes $q$ and a similar amplitude of variation in Stokes $u$. Figure~\ref{fig:stlouis_shock_slip_compare} shows a model that produces emission only from the two shock regions. This has strong phase-dependent polarization in both Stokes $q$ and $u$. Neither model produces polarization similar to the continuum signal. Both show the importance of considering line polarization and its capability to diagnose emission regions in colliding winds. It is clear that obtaining polarization observations specifically in spectral lines will provide additional information than that of the continuum polarization.


\section{Conclusions}

Colliding-wind binaries are important laboratories for the study of radiative wind driving, 
and massive stars in general, as the detailed structure and geometry of the interaction region between the winds can yield important information about each component's individual wind properties. This interaction is not only due to the bow shock, since the wind from each star can also be influenced by the radiation field from the other star. This phenomenon is best studied with a combination of spectroscopy and polarimetry, and as demonstrated above, the ultraviolet provides an ideal bandpass since the spectral energy densities of massive stars peak in that region, leading to a wavelength dependence of the polarization. 

Several analytical models provide a satisfactory description of the wind collision region, within certain simplifying assumptions detailed in Sect.~\ref{sec:pol}. Our group possesses the ability to leverage these models, as demonstrated for a few illustrative cases.

Using time series of spectropolarimetric observations from Polstar, we will be able to constrain the geometry of the bow shock and therefore the wind properties of a sample of 20 colliding-wind binaries spanning a large parameter space. The obtained dataset will significantly improve upon available observations; spectroscopy will be obtained with much higher signal-to-noise ratio and resolution than with IUE and WUPPE, and with greater polarimetric precision than the latter. 
Achieving a SNR of 100 at spectral resolution down to the wind sound speed,
over an orbitally resolved time domain, opens up new and unique wind dynamics diagnostics, as
the light from one star scatters in, and is absorbed by, the wind of the other, with orbitally
modulated changes in the relevant angles.
These produce a more precise mapping between radius and velocity than possible
for single stars, allowing tests of the degree to which different wind types and optical depths
alter the wind acceleration and structure formation.
Information about the bow shock also constrains the relative momentum fluxes in the winds
of the two different stars in the binary.


The work described in this white paper corresponds a single science objective included
in the \textit{Polstar} proposal, which overlaps in interesting ways
with several other objectives described in their own white papers. 
We will take advantage of the ways 
the unique diagnostics supplied by colliding wind binaries informs these other objectives.
For example, \citet{Gayley2021} discuss how \textit{Polstar} 
enables the study of structure and clumping in the winds of the $\sim 40$ brightest
targets, in
the context of single stars observed continuously over wind dynamical timescales.
Two of those targets, $\gamma$ Vel and $\delta$ Ori, are binary systems in the colliding-wind
target list, so our observations sampled over the system orbit will provide important
additional context for the continuous wind dynamics observations.
Also, \citet{Shultz2021} explore the global structural effects of
strong magnetization in a subset of OB stars, and a particularly important
example is Plaskett's star, which is the only strongly magnetic star observed to also
be a rapid rotator.
This star is also in our colliding wind target list, and the understanding of
its magnetic structure will be complemented by the methods made available by its short-period binary status,
as described here.

Closer binaries than the ones examined in this paper can interact and undergo episodes of 
conservative and nonconservative mass transfer, altering the mass and 
angular momentum of the binary constituents in ways
that are the subject of other \textit{Polstar} objectives and described 
in their own white papers. \citet{Peters2021} investigate how UV spectropolarimetry can characterize these mass flows and help determine the fraction of the mass loss by the system, whereas
\citet{Jones2021} focus on angular momentum evolution and the pathways
for stellar spinup and Keplerian disk creation.  UV spectropolarimetry can also provide information regarding the geometry, velocity, and inclination of such disks, as well as the accretion disks of Herbig Ae/Be stars \citep{Wisniewski2021}.

All of these objectives, whose realization will be made possible by 
\textit{Polstar}, constrain key evolutionary stages of massive stars and lead to a better understanding of their fates, including the compact objects that they leave behind. Importantly, the analysis of polarization signals from all of these sources will require a detailed understanding of the polarization induced by the intervening interstellar medium; by contributing to improve this knowledge, Polstar will not only enable these topics of stellar physics, but will also yield important insights in the physics of interstellar dust grains \citep{2021arXiv211108079A}.
Any of these types of systems can, in special cases, provide examples that can be studied
over an orbitally modulated range of angles that are either completely sampled, or
at least change considerably, over the timeframe of the \textit{Polstar} mission.
This white paper, therefore, is focused on the complementary spectropolarimetric
diagnostics available in 20 such cases.

\begin{acknowledgements}

RI acknowledges funding support from a grant by the National Science Foundation (NSF), AST-2009412. JLH acknowledges support from NSF under award AST-1816944 and from the University of Denver via a 2021 PROF award.
Scowen acknowledges his financial support by the NASA Goddard Space Flight Center to formulate the mission proposal for Polstar.
Y.N. acknowledges support from the Fonds National de la Recherche Scientifique (Belgium), the European Space Agency (ESA) and the Belgian Federal Science Policy Office (BELSPO) in the framework of the PRODEX Programme (contracts linked to XMM-Newton and Gaia).
NSL and CEJ wish to thank the National Sciences and Engineering Council of Canada (NSERC) for financial support.
A.D.-U. is supported by NASA under award number 80GSFC21M0002.
GJP gratefully acknowledges support from NASA grant 80NSSC18K0919 and STScI grants HST-GO-15659.002 and HST-GO-15869.001.  
\end{acknowledgements}

\bibliography{S5}

\end{document}